\documentclass[12pt]{article}
\pdfoutput=1
\usepackage{amsmath,amssymb,amsfonts,amsthm,bm,bbm,cancel,wasysym}
\usepackage{epsfig,graphics,graphicx,epstopdf,caption,subcaption}
\usepackage[numbers,sort&compress]{natbib}
\graphicspath{{Charts/}}
\usepackage{array,booktabs,colortbl,colordvi,multirow}
\usepackage{colordvi,color,xcolor}
\usepackage{hyperref}
\usepackage{rotating}
\usepackage{comment}
\usepackage{feynmf}

\usepackage[symbol]{footmisc}
\renewcommand{\thefootnote}{\fnsymbol{footnote}}
\newcommand{\eY}{\epsilon_{Y}}
\newcommand{\mY}{m_{Y}}
\begin{document}

\begin{center}

{\Large {\bf Self-interacting dark matter to freeze-in via vector portal}}\\

\vspace*{0.75cm}

{Xinyue Yin$^{a}$, Shuai Xu$^{b}$ and Sibo Zheng$^{a,}$\footnote{Corresponding author: sibozheng.zju@gmail.com.}}

\vspace{0.5cm}
{$^{a}$Department of Physics, Chongqing University, Chongqing 401331, China\\
$^{b}$School of Physics and Telecommunications Engineering, Zhoukou Normal University, Henan 466001, China}
\end{center}
\vspace{.5cm}

\begin{abstract}
\noindent
It is challenging to resolve the small-scale problem for dark matter being a weakly-interacting massive particle.
We attempt to address this issue by proposing a self-interacting freeze-in dark matter via Standard Model vector portal.
In this model, the dark matter obtains the observed relic abundance via $\gamma$ and $Z$ boson induced freeze-in processes, 
whereas the dark matter force mediator has a negligible relic abundance and a lifetime larger than the age of Universe.
We place constraints in classical and resonant regime resolving the small-scale problem from CMB, $X/\gamma$-ray, Supernova 1987A and out-of-equilibrium condition. 
It turns out that the CMB constraint on dark matter annihilations is satisfied despite large Sommerfeld effect taking place, 
while the other constraints are trivially accommodated due to various millicharge induced suppressions. 
Finally we briefly discuss future cosmological tests on such freeze-in dark matter model. 
\end{abstract}

\renewcommand{\thefootnote}{\arabic{footnote}}
\setcounter{footnote}{0}
\thispagestyle{empty}
\vfill
\newpage
\setcounter{page}{1}

\tableofcontents

\section{Introduction}
It is known that dark matter (DM) is the major matter content of Universe.
DM as a weakly interacting massive particle (WIMP) has been searched for in a number of direct detection experiments.
However, no solid signal has been reported so far.
This initiates a renewed interest in alternative DM scenarios such as freeze-in DM (FIDM), see e.g, \cite{Bernal:2017kxu} for a review.
Unlike WIMP-like DM, a FIDM is instead produced in the early Universe via so-called freeze-in mechanism \cite{Hall:2009bx} 
due to a feeble interaction between the Standard Model (SM) and DM.
This feeble interaction makes the FIDM be unable to keep thermal equilibrium with the SM thermal bath in the early Universe.

Apart from coupling to the SM sector, DM (labeled by $X$ hereafter) may also self-interact with each other.
As firstly proposed by \cite{Spergel:1999mh} and verified by subsequent N-body simulations \cite{Rocha:2012jg,Peter:2012jh,Zavala:2012us,Elbert:2014bma},
self-interacting DM can address the small-scale problem related to observations of small-scale structure from dwarf to cluster scales, by providing a DM scattering cross section
$\sigma_{T}/m_{X}\sim 1.0~\rm{cm}^{2}/\rm{g}$ \cite{Dave:2000ar,1201.5892,1208.3025,1412.1477,Oh:2010ea,KuziodeNaray:2007qi} with $v\sim10$ km/s 
and $\sigma_{T}/m_{X}\sim 0.1-1.0~\rm{cm}^{2}/\rm{g}$ \cite{Kaplinghat:2015aga,Robertson:2016xjh,Elbert:2016dbb,Bondarenko:2017rfu,Harvey:2018uwf,Sagunski:2020spe,Newman:2015kzv, Newman:2012nw,Newman:2012nv} with $v\sim10^{3}$ km/s,
where $\sigma_T$ is the DM scattering cross section, $m_X$ the DM mass and $v$ the DM relative velocity respectively.
See \cite{Tulin:2017ara} for a review on this topic.
The mild velocity-dependence of $\sigma_T$ on the DM relative velocity can be interpreted as a result of Yukawa potential \cite{Buckley:2009in,Tulin:2012wi,Tulin:2013teo,Colquhoun:2020adl} due to a DM force mediator $Y$ with mass $m_{Y}$ far lighter than $m_{X}$.

In this work we consider self-interacting FIDM as a solution to the small-scale problem. 
Obviously, the FIDM is easier to evade current DM direct and indirect detection limits, 
compared to the WIMP-like DM based on freeze-out mechanism. 
To address the small-scale problem, the FIDM sector should at least include the aforementioned DM and the DM force mediator. 
In the context of two-field FIDM, there are two types of arrangements on the feeble interaction between the SM and FIDM sector: 
(i) one assumes $Y$ coupled to the SM sector with strong coupling strength and $X$ coupled to $Y$ with coupling strength being feeble respectively; and (ii) one adopts $X$ coupled to the SM sector with feeble coupling constant and $Y$ coupled to $X$ with strong coupling strength respectively. 
The first situation is LHC-friendly as studied by \cite{Hessler:2016kwm,Ghosh:2017vhe,Calibbi:2018fqf,Belanger:2018sti},
while the second situation is able to resolve the small-scale problem. 

The rest of this paper is organized as follows.
In Sec.\ref{model}, we firstly present the self-interacting FIDM via SM vector portal with an emphasize on model parameters.
Then we use current small-scale data to identify both classical and resonant regime resolving the small-scale problem.
Afterward, we calculate the DM freeze-in relic abundance to fix the remaining model parameter - the DM millicharge,
where we point out a key difference between this work and the conventional millicharged DM. 
Sec.\ref{cons} is devoted to discuss various constraints in the classical or resonant regime, 
including CMB constraints on DM annihilations into SM particles and DM scattering off baryons where Sommerfeld effects \cite{Sommerfeld} take place in the former situation, 
$X/\gamma$-ray constraint on the DM force mediator as it has a lifetime bigger than the age of Universe,
Supernova 1987A constraint on the DM force mediator, and out-of-equilibrium constraint on the entire DM sector.
We will verify that all of the constraints considered are satisfied.
Finally we conclude in Sec.\ref{con}, with a brief discussion about future cosmological tests on this FIDM model.


\section{The two-field dark matter sector}
\label{model}
We consider a two-field FIDM sector with the following Lagrangian 
\begin{eqnarray}\label{Lag1}
\mathcal{L}_{\rm{dark}}&=&\bar{X}\left(i\gamma^{\nu}\partial_{\nu}-\mu_{X}\right)X+\frac{1}{2}\partial_{\mu}Y\partial^{\mu}Y-\frac{\mu^{2}_{Y}}{2}Y^{2}-\lambda_{D}\bar{X}XY-\frac{\lambda_{Y}}{4}(Y^{2}-\upsilon^{2}_{Y})^{2}\nonumber\\
&-&\epsilon' e \bar{X}\gamma^{\mu}XB_{\mu},
\end{eqnarray}
where  $B_{\mu}$ is the SM $U(1)_Y$ gauge field, $X$ is the DM fermion carrying a millicharge $\epsilon'$ \footnote{The millicharge term can be produced in various contexts such as  kinetic mixing \cite{Holdom:1985ag, Feldman:2007wj}.} of $U(1)_Y$ in units of $e$,
real scalar $Y$ is the DM force mediator, and $\mu_{X}$, $\mu_{Y}$ are bare masses. 
To be in accord with our setup of the portal interaction being $X$-SM type, 
terms of $Y$-SM type such as $Y^{2}\mid H\mid^{2}$ are assumed to be much smaller. 

In the broken electroweak phase with $B_{\mu}=-\sin\theta_{W}Z_{\mu}+\cos\theta_{W}A_{\mu}$, where $\theta_{W}$ is the weak mixing angle, $Z_{\mu}$ the $Z$ boson and $A_{\mu}$ the photon,
 the DM Lagrangian in eq.(\ref{Lag1}) can be rewritten as 
\begin{eqnarray}\label{Lag3}
\mathcal{L}^{\rm{eff}}_{\rm{dark}}&=&\bar{X}\left(i\gamma^{\nu}\partial_{\nu}-\mu_{X}\right)X+\frac{1}{2}\partial_{\mu}Y\partial^{\mu}Y-\frac{\mu^{2}_{Y}}{2}Y^{2}-\lambda_{D}\bar{X}XY-\frac{\lambda_{Y}}{4}(Y^{2}-\upsilon^{2}_{Y})^{2},\nonumber\\
&-&\epsilon e\bar{X}\gamma^{\mu}\left(-\tan\theta_{W}Z_{\mu}+A_{\mu}\right)X,
\end{eqnarray}
where $\epsilon =\epsilon'\cos\theta_{W}$.
Expanding the field $Y$ around the vacuum expectation value $\upsilon_{Y}$ in eq.(\ref{Lag3}), one obtains the final form of the effective Lagrangian
\begin{eqnarray}\label{Lag4}
\mathcal{L}^{\rm{eff}}_{\rm{dark}}&=&\bar{X}\left(i\gamma^{\nu}\partial_{\nu}-m_{X}\right)X+\frac{1}{2}\partial_{\mu}Y\partial^{\mu}Y-\frac{m^{2}_{Y}}{2}Y^{2}\nonumber\\
&-&\lambda_{D}\bar{X}XY-\epsilon e\bar{X}\gamma^{\mu}\left(-\tan\theta_{W}Z_{\mu}+A_{\mu}\right)X,
\end{eqnarray}
with $m_{X}=\mu_{X}+\lambda_{D}\upsilon_{Y}$ and $m^{2}_{Y}=\mu^{2}_{Y}+2\lambda_{Y}\upsilon^{2}_{Y}$, where $Y$ self-interaction terms have been ignored.

The effective Lagrangian in eq.(\ref{Lag4}) shows that $X$ can be understood as a generalized millicharged DM \cite{Feldman:2007wj, Davidson:2000hf,Chuzhoy:2008zy,McDermott:2010pa,Dolgov:2013una,Dvorkin:2013cea}, 
as it additionally couples to the $Z$ boson and $Y$ scalar singlet in this model. 
The $Z$ coupling contributes to new DM freeze-in processes in the DM mass range of $m_{X}<m_{Z}/2$, 
whereas the $Y$ coupling with $\lambda_{D}$ being of order unity makes the DM strongly self-interact with each other. 

To make phenomenological analysis be robust, 
we assume the bare masses to be negligible for simplicity, 
and focus on the set of model parameters $\{\upsilon_{Y}, \lambda_{Y}, \lambda_{D}, m_{X}, m_{Y}, \epsilon\}$.
Consider that the first two can be eliminated by the mass relations mentioned above,
we choose the following independent variables
\begin{eqnarray}\label{pars}
\{\lambda_{D}, m_{X},m_{Y}, \epsilon\},
\end{eqnarray}
where the first three are related to the DM self-interaction while the last two determine the DM freeze-in production, respectively.

\subsection{Dark matter self-interaction}
\label{sin}
In this section we consider the non-relativistic DM scattering due to a Yukawa potential
\begin{eqnarray}\label{potential}
V(r)=-\frac{\alpha_{D}}{r}e^{-m_{Y}r},
\end{eqnarray}
where $\alpha_{D}=\lambda^{2}_{D}/4\pi$ is the dark fine structure constant.
The main task of this section is to identify parameter regions as composed of $\{\alpha_{D}, m_{X}, m_{Y}\}$ where 
the small-scale data in Table \ref{bv} is satisfied. 

In Table \ref{bv} the averaged cross section is defined as 
\begin{equation}\label{asigma}
\bar{\sigma} = \int\sigma_{T}(v)f(v)dv,
\end{equation}
where $f(v)$ is the Maxwell-Boltzmann distribution function with dispersion velocity $v_{0}\approx220$ km/s.
Under various limits \cite{Tulin:2013teo} such as Born, classical and resonant regime,
$\sigma_T$ in eq.(\ref{asigma}) can be approximately written by analytical formulae.
In the Born region with respect to $\alpha_{D}m_{X}<<m_{Y}$, the cross section $\sigma^{\rm{Born}}_{T}$ is not sensitive to $v$.
In the following we only focus on the classical and resonant regime.

\begin{table}
\begin{center}
\begin{tabular}{cccc}
\hline\hline
System & $\left<v\right>$~& $\bar{\sigma}\left<v\right>/m_{X}$ &Refs.\\ \hline
dwarf & 50 ~&  [300,700] & \cite{Zavala:2012us,Elbert:2014bma} \\
galaxy & 250 &  [400,1000]& \cite{Zavala:2012us,Elbert:2014bma} \\
group & 1150 & [345,805]&\cite{Sagunski:2020spe,Newman:2015kzv}\\
cluster &1900  & [190,570]&\cite{Sagunski:2020spe,Newman:2012nw,Newman:2012nv}\\
\hline \hline
\end{tabular}
\caption{The observed and favored values of $\bar{\sigma}\left<v\right>/m_{X}$ in units of (cm$^2$/g)(km/s) at group to cluster scale with $\left<v\right>\sim 10^{3}$ km/s and dwarf to galaxy scale $\left<v\right>\sim 10^{2}$ km/s, respectively,
where $\left<v\right>$ is the averaged velocity.}
\label{bv}
\end{center}
\end{table}

\subsubsection{Classical regime}
In the classical regime with respect to $m_{X}v>>m_{Y}$, the DM scattering cross section $\sigma_{T}$ in eq.(\ref{asigma}) is given by \cite{Colquhoun:2020adl}
\begin{equation}{\label{sigma}}
\sigma^{\rm{clas}}_{T}\approx
 \left\{
\begin{array}{lcl}
\frac{2\pi}{m^{2}_{Y}}\beta^{2}\rm{ln}(1+\beta^{-2}), ~~~~~~~~~~~~~~~~~~~~~~~~~~~~\beta\leq 10^{-2},\\
\frac{7\pi}{m^{2}_{Y}}\frac{\beta^{1.8}+280(\beta/10)^{10.3}}{1+1.4\beta+0.006\beta^{4}+160(\beta/10)^{10}}, ~~~~~~~10^{-2}\leq \beta\leq 10^{2},\\
\frac{0.81\pi}{m^{2}_{Y}}\left(1+\rm{ln}\beta-(2\rm{ln}\beta)^{-1}\right)^{2}, ~~~~~~~~~~~~~~~~\beta\geq 10^{2},\\
\end{array}\right.
\end{equation}
for the potential in eq.(\ref{potential}),
where $\beta=2\alpha_{D}m_{Y}/(m_{X}v^{2})$. Eq.(\ref{sigma}) shows that $\sigma^{\rm{clas}}_{T}\sim v^{-4}$ and $\sigma^{\rm{clas}}_{T}\sim v^{-0.6}$ for $\beta\leq 10^{-2}$ and $10\leq\beta\leq 10^{2}$ respectively.
The dependences of $\sigma^{\rm{clas}}_{T}$ on $v$ in these regions are favored by small-scale data shown in Table \ref{bv}.  
 
\begin{figure}
\centering
\includegraphics[width=8cm,height=8cm]{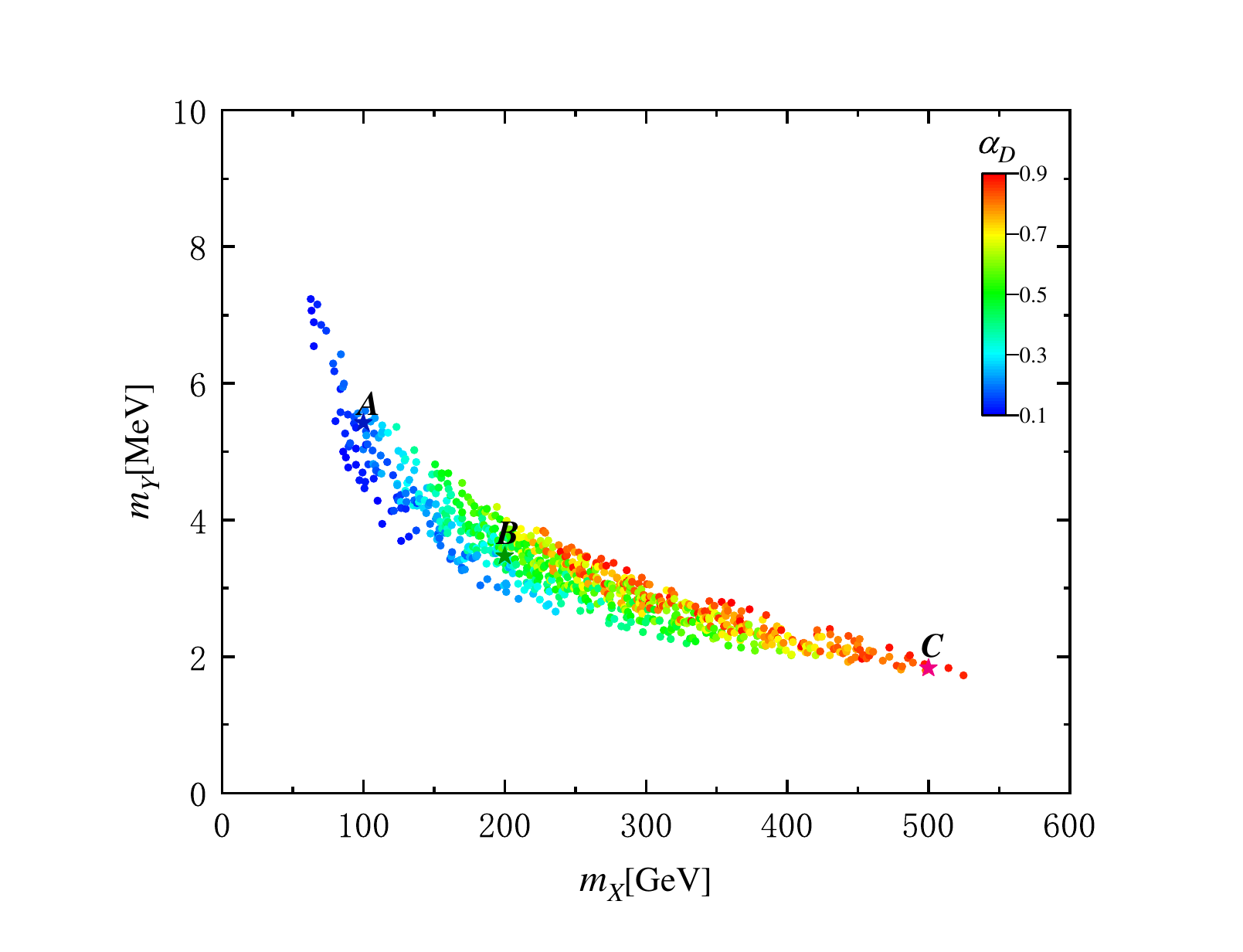}
\includegraphics[width=8cm,height=8cm]{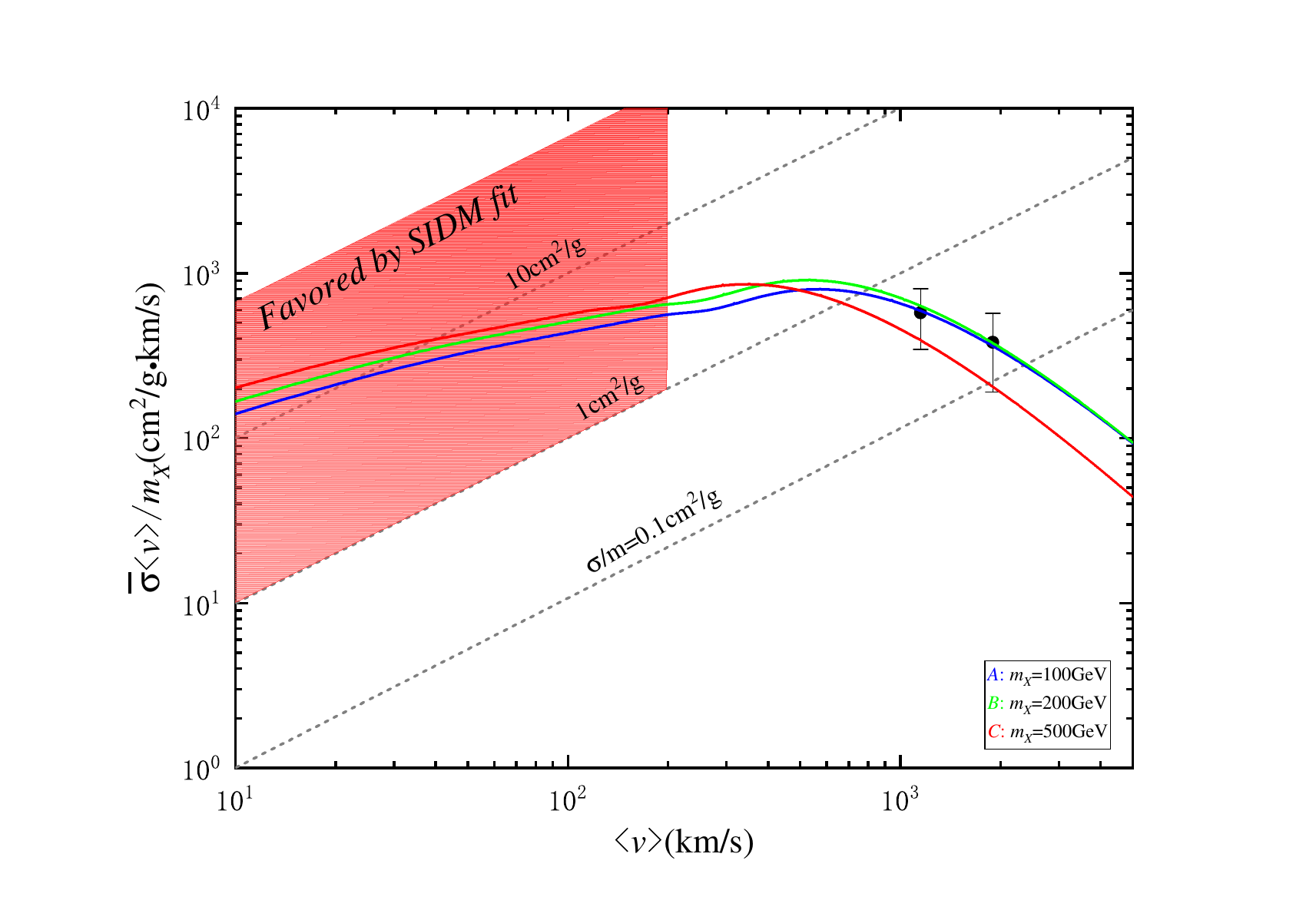}
\centering
\caption{Left: the classical regime resolving the small-scale data in Table.\ref{bv}. 
Right: three benchmark points $A$, $B$ and $C$ in the left panel are projected to the plane of $(\bar{\sigma}\left<v\right>/m_{X})-\left<v\right>$ for a comparison with the small-scale observations. See text for details.}
\label{Classf}
\end{figure}

 Substituting eq.(\ref{sigma}) into eq.(\ref{asigma}), 
 we show in fig.\ref{Classf} the samples (left) where  
both the small-scale data in Table \ref{bv} and the condition $m_{X}v>>m_{Y}$ are simultaneously satisfied,
 and three explicit benchmark points $A$, $B$ and $C$ (right) for illustration. 
In the left panel, the set of these samples, referred to as the classical regime, include the previously reported ones in \cite{Colquhoun:2020adl}.
In the right panel, the benchmark points  $A$, $B$, and $C$ correspond to the explicit values of $\alpha_{D}=\{0.16, 0.43, 0.89\}$, $m_{Y}=\{5.4, 3.4, 1.8\}$ MeV and $m_{X}=\{100, 200, 500\}$ GeV, respectively.

\begin{figure}
\centering
\includegraphics[width=8cm,height=8cm]{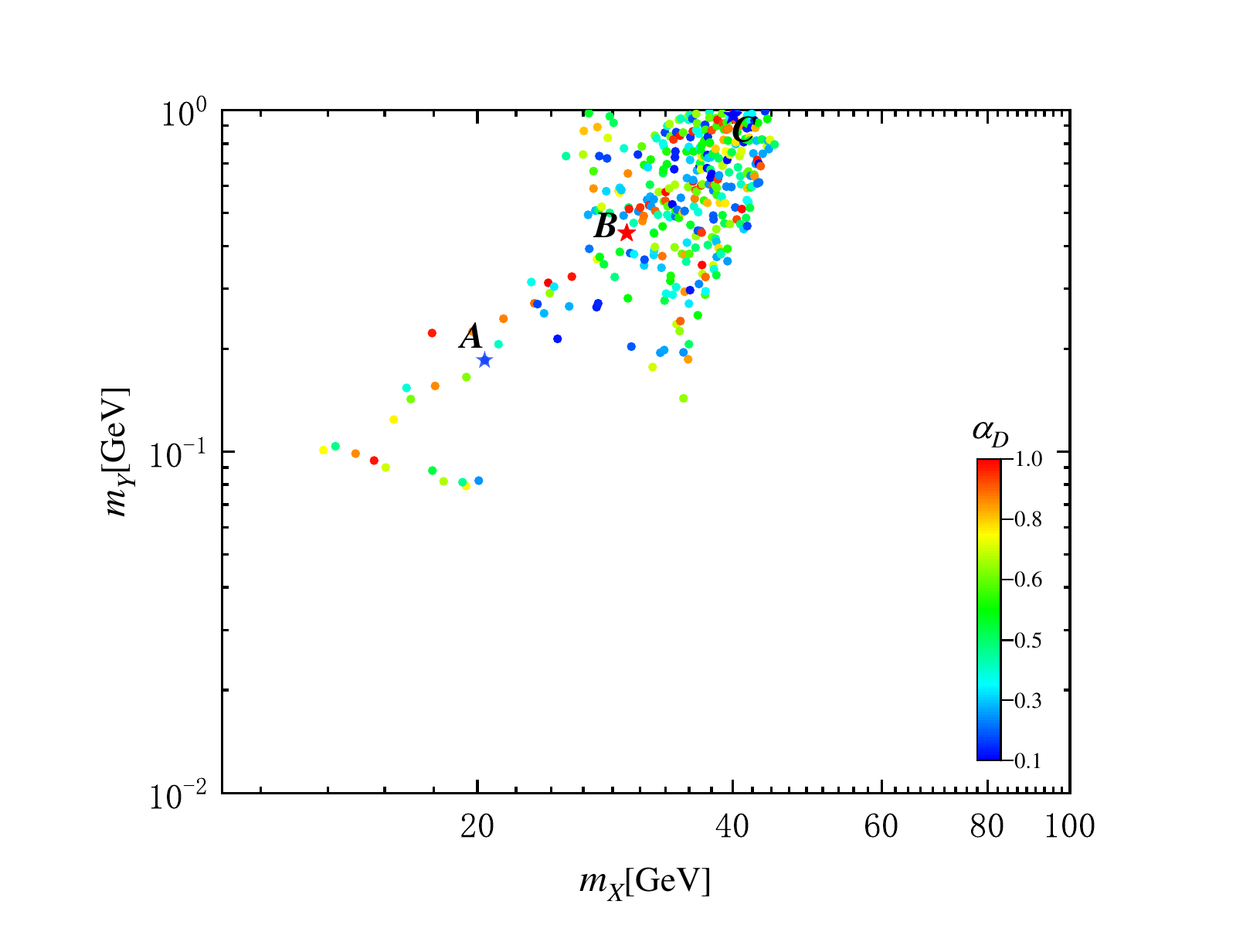}
\includegraphics[width=8cm,height=8cm]{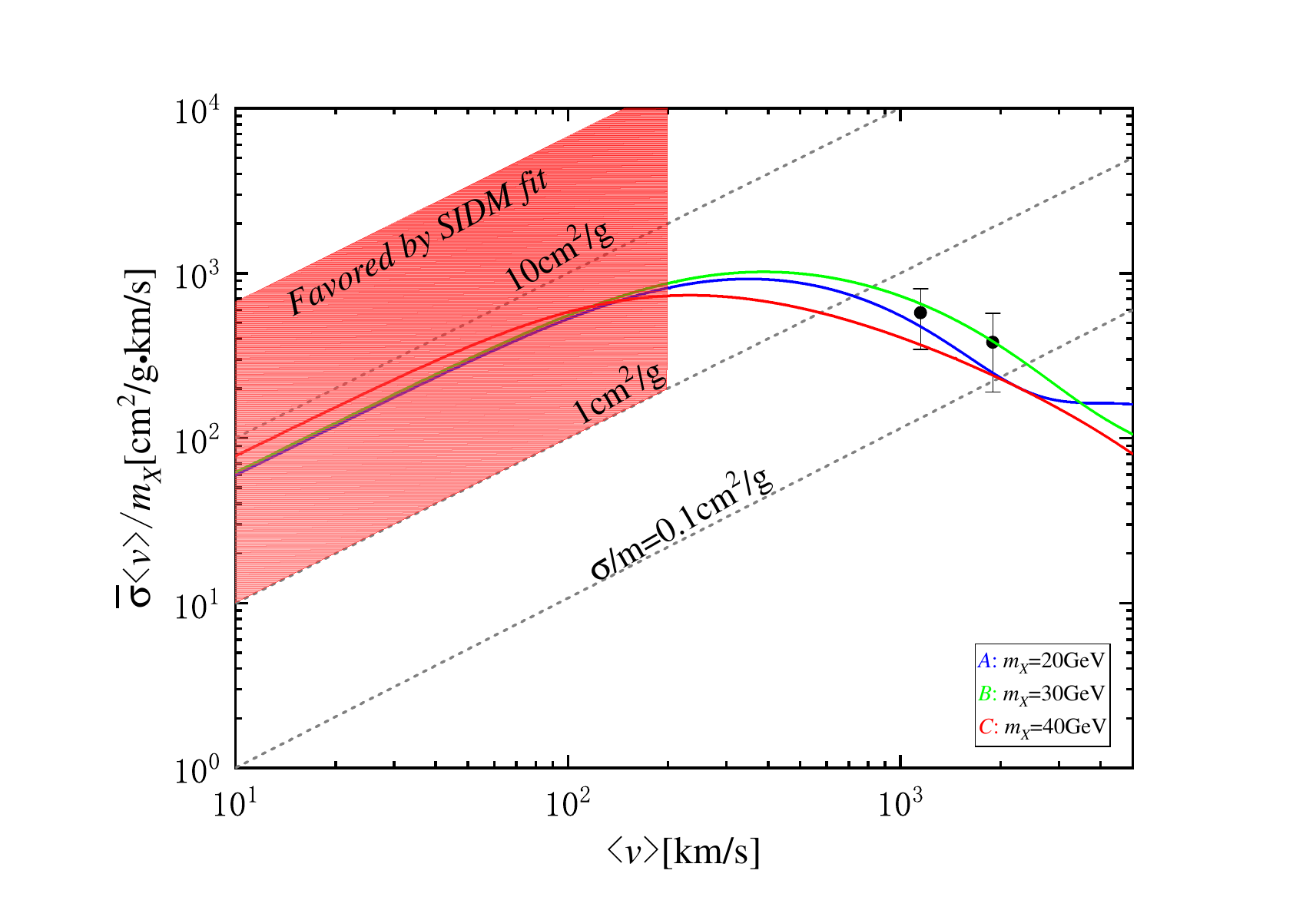}
\centering
\caption{Same as fig.\ref{Classf} but in the resonant regime, where the values of $m_X$ are in a more narrow mass range.}
\label{Resf}
\end{figure}

\subsubsection{Resonant regime}
In the resonant regime with respect to $m_{X}v\leq m_{Y}$ and $\alpha_{D}m_{X}\geq m_{Y}$, the DM scattering cross section is obtained either via solving the Schrodinger equation in terms of a partial wave analysis, 
or described by an analytic approximation for the s-wave scattering cross section being valid in certain resonant regime. 
Here it is sufficient to adopt the latter one \cite{Tulin:2013teo}
\begin{equation}{\label{sigmar}}
\sigma^{\rm{res}}_{T}\approx \frac{16\pi}{m^{2}_{X}v^{2}}\sin^{2}(\delta)
\end{equation}
where
\begin{equation}{\label{delta}}
\delta=\arg\left(\frac{i\Gamma\left(\frac{im_{X}v}{\kappa m_{Y}}\right)}{\Gamma(\lambda_{+})\Gamma(\lambda_{-})}\right)
\end{equation}
with 
\begin{equation}{\label{lambda}}
\lambda_{\pm}=1+\frac{im_{X}v}{2\kappa m_{Y}}\pm\sqrt{\frac{\alpha_{D}m_{X}}{\kappa m_{Y}}-\frac{m^{2}_{X}v^{2}}{4\kappa^{2} m^{2}_{Y}}},
\end{equation}
and $\kappa\approx 1.6$.

Fig.\ref{Resf} shows the samples where the small-scale data in Table \ref{bv} and the conditions of resonant regime $m_{X}v\leq m_{Y}$ and $\alpha_{D}m_{X}\geq m_{Y}$ are satisfied simultaneously, 
and three benchmark points (right) for illustration.
In the left panel, the validity of $\sigma^{\rm{res}}_{T}$ in eq.(\ref{sigmar}) for each sample has been checked by  using the numerical analysis in \cite{Tulin:2013teo}. We will refer to the set of these samples as the resonant regime.
In the right panel, the three benchmark points $A$, $B$, $C$ correspond to the explicit values of $\alpha_{D}=\{0.23, 0.85, 0.15\}$, $m_{Y}=\{0.18, 0.44, 0.98\}$ GeV and $m_{X}=\{20, 30, 40\}$ GeV respectively.
Compared to the classical regime as shown in fig.\ref{Classf}, the resonant regime in fig.\ref{Resf} asks for a relatively larger $m_Y$ and a more narrow mass range of $m_X$.

\begin{figure}
\centering
\includegraphics[width=12cm,height=8cm]{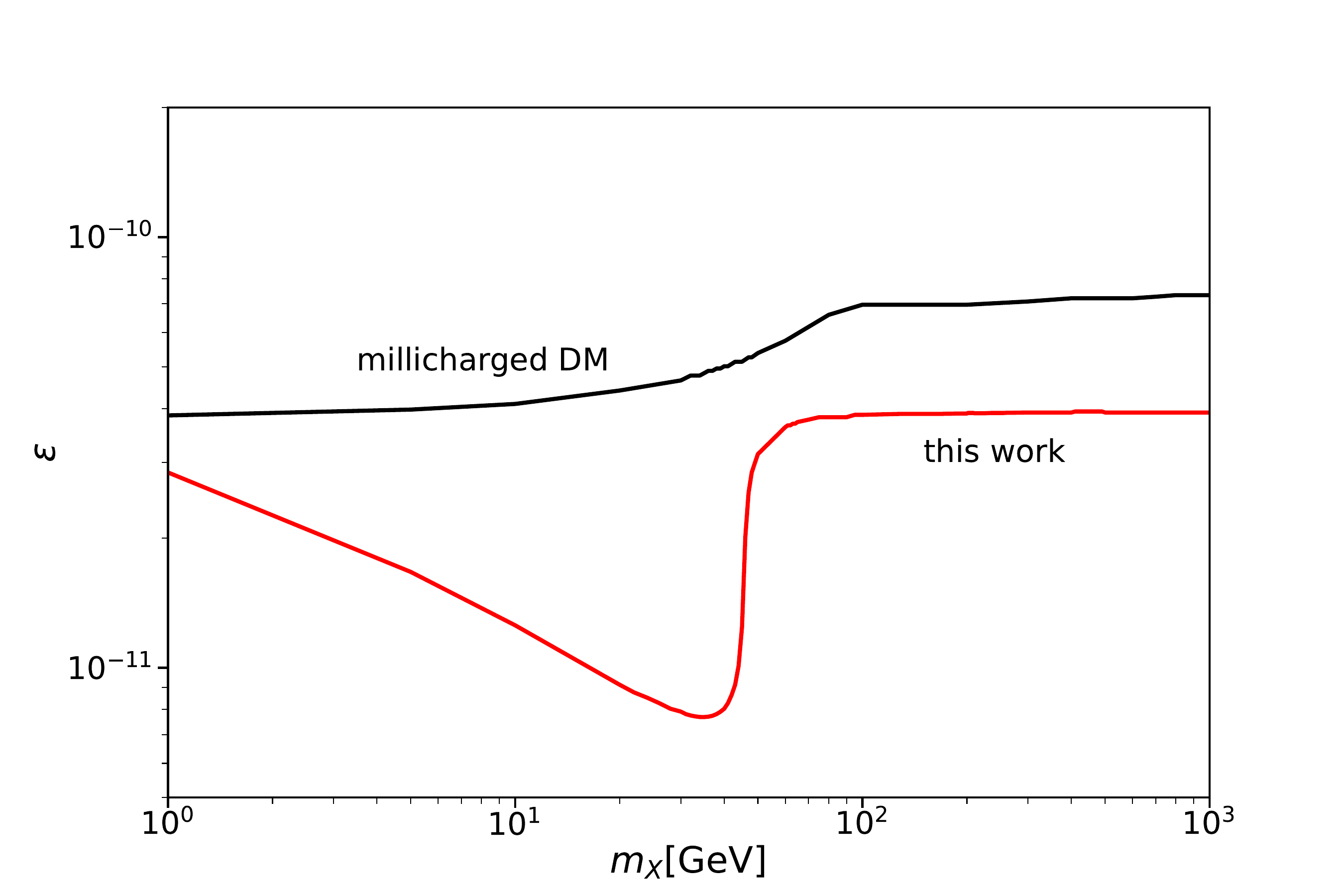}
\centering
\caption{The observed DM relic density projected to the two-dimensional plane of $\epsilon-m_{X}$ in this work (red), 
compared to the millicharged DM (black).}
\label{relic}
\end{figure}

\subsection{Freeze-in relic abundance}
\label{rel}
The DM is produced via the SM $Z$ and/or $\gamma$ induced freeze-in processes.
Using the publicly available code micrOMEGAs5.0 \cite{Belanger:2018ccd},
we show in fig.\ref{relic} the observed DM relic abundance $\Omega_{X}h^{2}=0.12\pm0.001$ \cite{Planck:2018vyg} projected to the two-dimensional plane of $m_{X}$ and $\epsilon$, where our model (in red) is compared to the millicharged DM (in black). 
Three remarks are in order regarding this result. 
\begin{itemize}
\item First, given an explicit value of $m_X$ the required value of $\epsilon$ in our model is a few times smaller than in the millicharged DM,
as a result of the coupling of DM to $Z$ boson in eq.(\ref{Lag4}). 
While safely neglected in low energy regions, 
this coupling contributes to additional DM freeze-in channels.
For example, the decay channel $Z\rightarrow X\bar{X}$ dominates the DM production in the DM mass range of $m_{X}<m_{Z}/2$,
showing why large differences occur in this DM mass region.
\item Second, there is a dramatical change in the curve of DM relic abundance at $m_{X}\approx m_{Z}/2$. 
The reason is that (i) in the DM mass range of $m_{X}<m_{Z}/2$ the DM production is dominated by the $Z$ boson decay $Z\rightarrow X\bar{X}$ as mentioned above,
whereas (ii) in the DM mass range of $m_{X}>m_{Z}/2$ this channel is kinetically forbidden and the DM production is instead dominated by the 
SM particle annihilations $f\bar{f}\rightarrow X\bar{X}$, with $f$ being SM fermions.
\item Third,  we have neglected Sommerfeld effects on the DM freeze-in production. 
This is justified by the study of \cite{Du:2021jcj} on SM annihilation induced FIDM phase-space distribution as in our case,
which shows the DM averaged momentum $\left<p_{X}\right>\geq m_{X}$ during the DM freeze-in process, 
suggesting the averaged DM relative velocity $\left<v\right>\approx 1$.
So, the Sommerfeld effects are negligible.
\end{itemize}

\begin{figure}
\centering
\includegraphics[width=8cm,height=8cm]{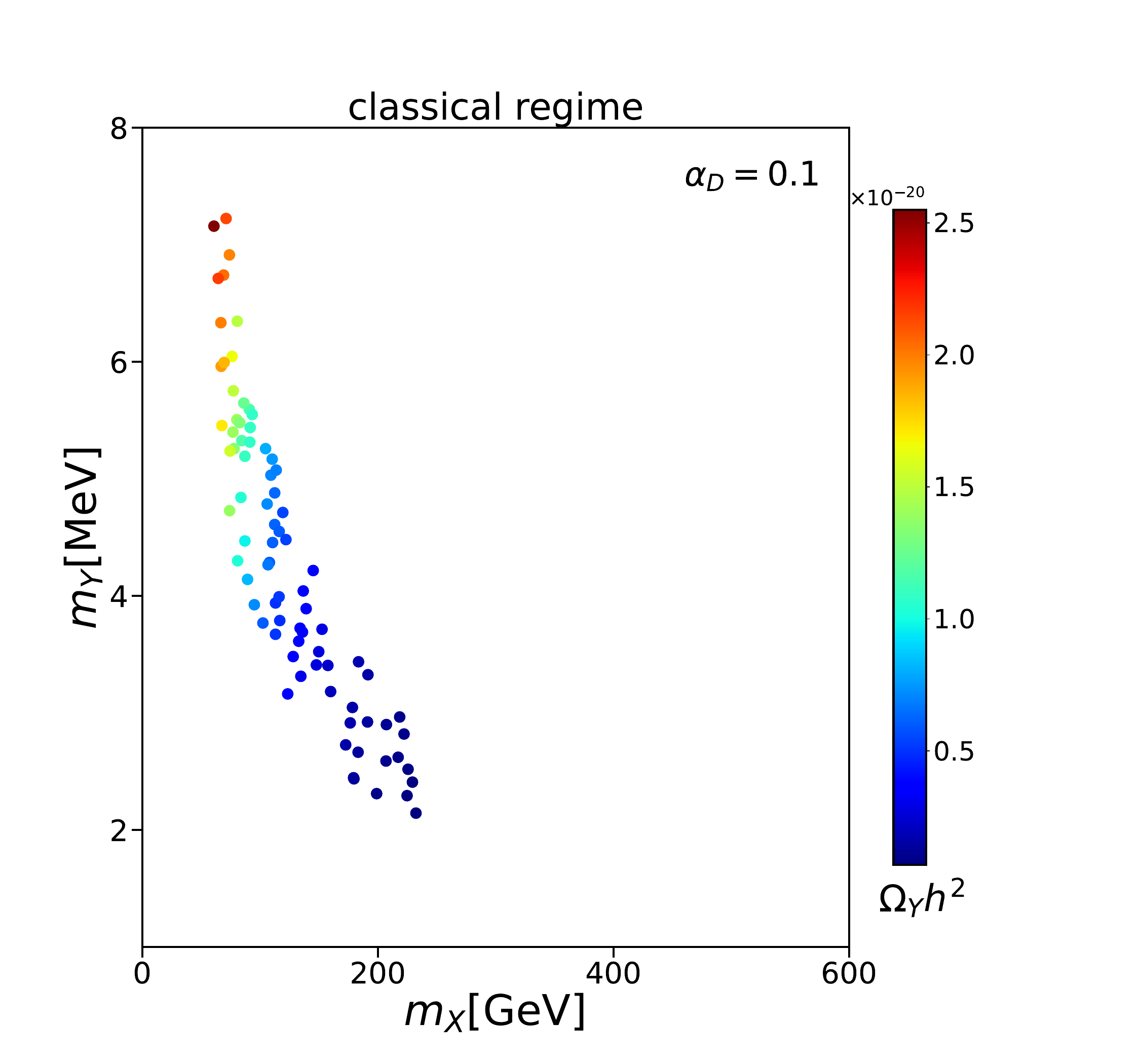}
\includegraphics[width=8cm,height=8cm]{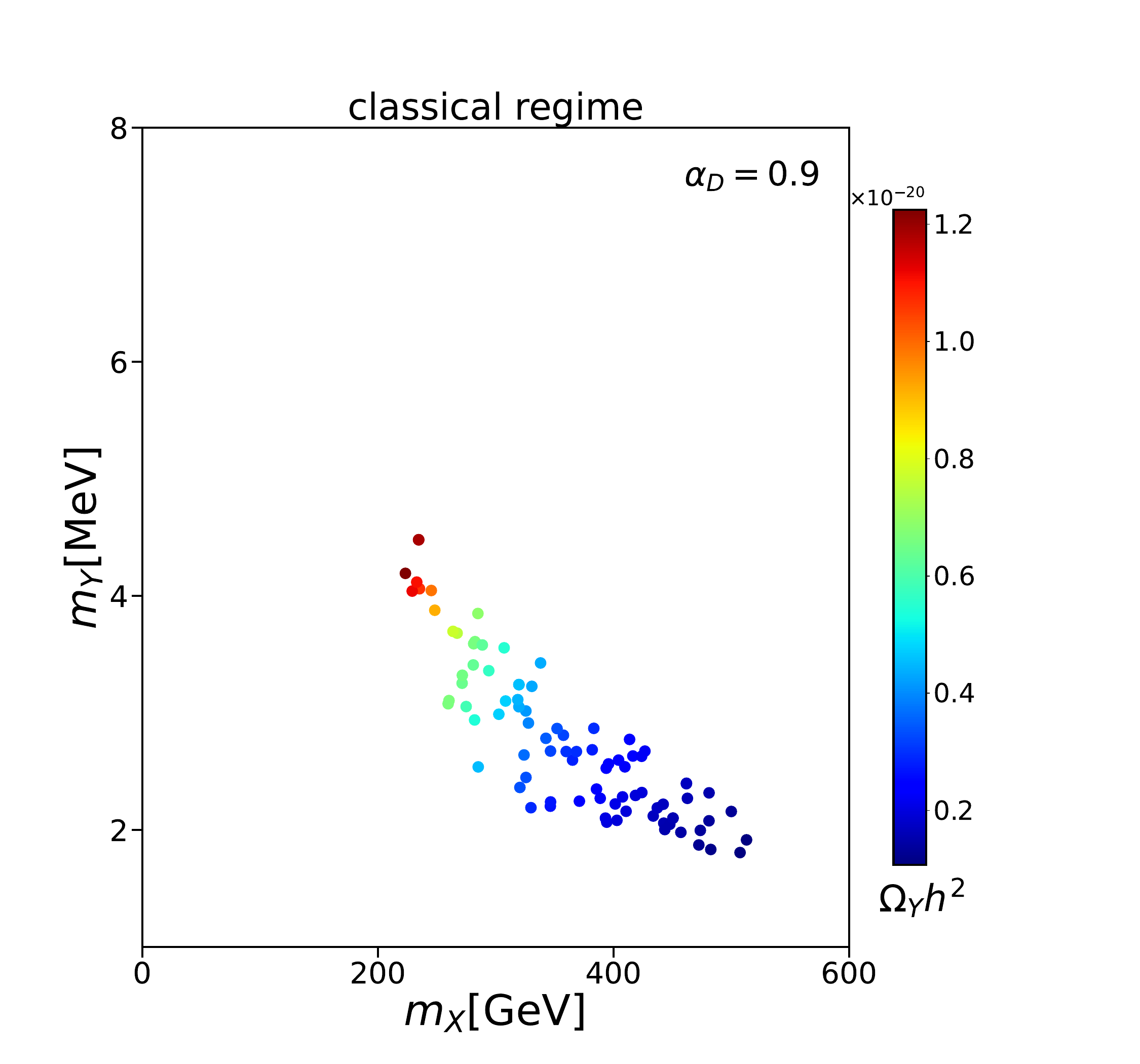}\\
\includegraphics[width=8cm,height=8cm]{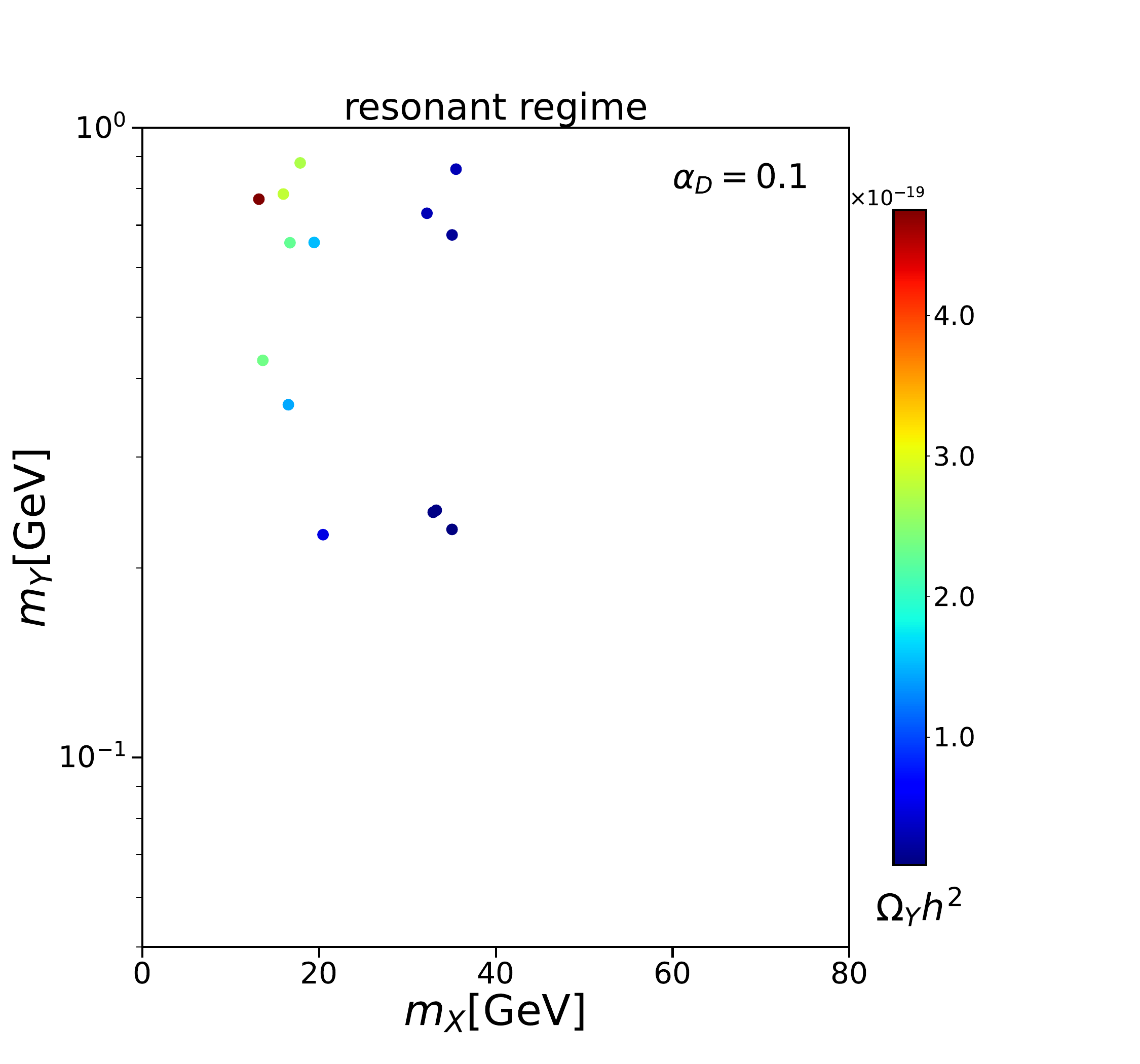}
\includegraphics[width=8cm,height=8cm]{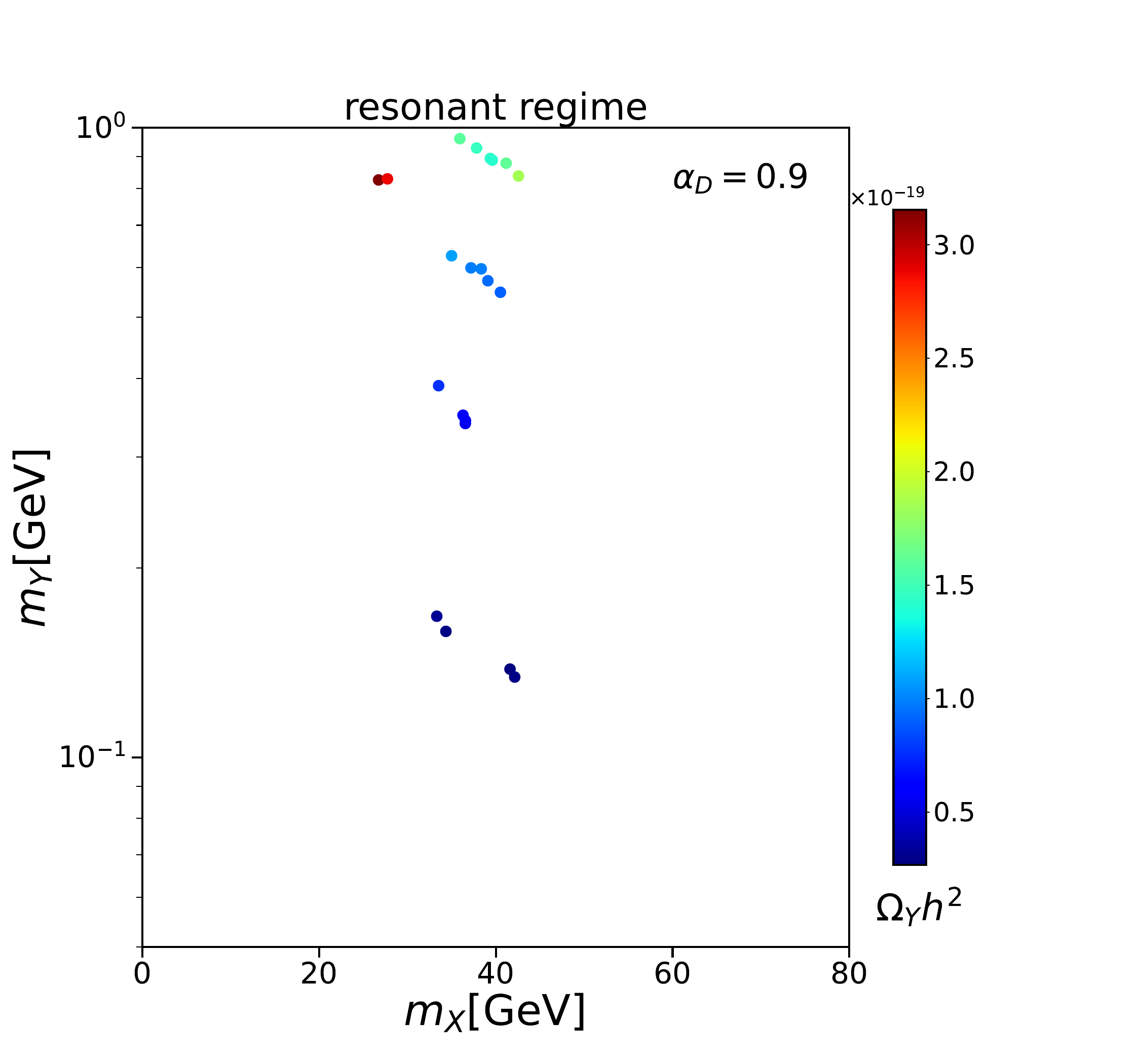}
\centering
\caption{Contours of $\Omega_{Y}h^{2}$ in the classical (top) and resonant (bottom) regime for $\lambda_{D}=0.1$ (left) and $0.9$ (right) extracted from figs.\ref{Classf}-\ref{Resf} respectively, 
where the values of millicharge $\epsilon$ are fixed by the observed DM relic density shown in fig.\ref{relic}.}
\label{Yrelic}
\end{figure}

Additionally, the DM force mediator relic density $\Omega_{Y}h^{2}$ has to be considered if its lifetime $\tau_{Y}$ is larger than the age of Universe, verified by fig.\ref{lifetime} as below.  
To derive $\Omega_{Y}h^{2}$, we employ an effective field theory analysis.
While being neutral, the singlet scalar $Y$ effectively couples to the SM sector due to DM loop as\footnote{The DM loop also yields a coupling of $Y$ to $Z$ bosons with dimension 5. This coupling induced inverse decay of $Y\rightarrow ZZ^{*}$ is however prohibited by $m_{Y}<m_{Z}$ considered here.
Therefore, the inverse decay of $Y\rightarrow \gamma\gamma$ due to eq.(\ref{eff}) is the main freeze-in process.
Having said so, the scattering $\gamma q\rightarrow Y q$ with $q$ the SM fermions may take over the inverse decay for a  reheating temperature high enough. Here, we simply choose a low reheating temperature.}
\begin{equation}{\label{eff}}
\mathcal{L}_{Y,\rm{eff}} = c_{\gamma}YF_{\mu\nu}^{2},
\end{equation}
with
\begin{eqnarray}{\label{cs}}
c_{\gamma} &=&\frac{\epsilon^{2}\alpha\lambda_{D}}{2\pi m_{X}},
\end{eqnarray}
where $\alpha$ is the fine structure constant. Note that the $c_{\gamma}$ coupling differs from axion-like coupling to photons. 
According to eq.(\ref{eff}),  we show in fig.\ref{Yrelic}  the contours of $\Omega_{Y}h^{2}$ in the classical (top) and resonant (bottom) regime 
addressing the small-scale problem for $\alpha_{D}=0.1$ (left) and 0.9 (right) extracted from figs.\ref{Classf}-\ref{Resf} respectively.
This figure shows small values of $\Omega_{Y}h^{2}$ ranging from $\sim 10^{-21} (10^{-20})$ to $\sim 10^{-20} (10^{-19})$ in the classical (resonant) regime, 
as a result of $\epsilon^{4}$-suppression on $Y$ freeze-in rate.

A relatively small value of $\Omega_{Y}h^{2}$, compared to $\Omega_{X}h^{2}$, 
hardly affects the one-to-one correspondence between $\epsilon$ and $m_{X}$ in fig.\ref{relic}.
In what follows we will use fig.\ref{relic} to estimate various constraints in the classical regime shown in fig.\ref{Classf} or the resonant regime shown in fig.\ref{Resf}.

\section{Constraints}
\label{cons}
In this section we discuss constraints relevant in the classical or resonant regime.
\subsection{CMB}
\subsubsection{Dark matter annihilations}
DM annihilations into electrically charged particles distort CMB spectra.
Planck data has been used to set the CMB constraint \cite{Slatyer:2015jla} on DM annihilation cross sections,
which will be referred to as the CMB (ann) constraint.
In our situation the DM annihilation cross section is \cite{McDermott:2010pa}
\begin{equation}{\label{ann}}
\left<\sigma_{\rm{ann}} v\right>_{\bar{f}f} =(\sigma_{\rm{ann}}v)_{\bar{f}f} \frac{x^{3/2}}{2\sqrt{\pi}}\int_{0}^{\infty} S(v)v^{2}e^{-xv^{2}/4}dv,
\end{equation}
by assuming the Maxwell-Boltzmann distribution, 
where 
\begin{equation}{\label{anntree}}
(\sigma_{\rm{ann}}v)_{\bar{f}f}=\frac{\pi\alpha\epsilon^{2}}{m^{2}_{X}}q^{2}_{f}N_{c}\sqrt{1-\frac{m^{2}_{f}}{m^{2}_{X}}}\left(1+2\frac{m^{2}_{f}}{m^{2}_{X}}\right)
\end{equation}
with $f$ the SM final states, $m_f$ and $q_{f}$ the mass and electric charge of $f$ respectively, $N_{c}$ the color factor, i.e, $N_{c}=3~(1)$ for quarks (leptons), $x\approx v^{-2}_{0}$, 
and the Sommerfeld factor \cite{Feng:2010zp}\footnote{To obtain the Sommerfeld factor, 
one either solves the non-relativistic schrodinger equation with the Yukawa potential in eq.(\ref{potential}), 
or uses analytical formulae \cite{Iengo:2009ni,Cassel:2009wt,Slatyer:2009vg,Feng:2010zp}.
Here, it is sufficient to use eq.(\ref{sapp}).} 
\begin{eqnarray}\label{sapp}
S(v)\approx  \frac{\pi}{\epsilon_v}
\frac{\sinh \left( \frac{2\pi\epsilon_v}{\pi^2\eY/6}
  \right)}
{ \cosh \left( \frac{2\pi\epsilon_v} {\pi^2 \eY / 6} \right)
- \cos \left( 2\pi\sqrt{\frac{1}{\pi^2\eY/6}
-\frac{\epsilon^2_v}{(\pi^2\eY/6)^2}} \, \right) } \ ,
\end{eqnarray}
with $\epsilon_v \equiv v / \alpha_D$ and $\eY \equiv \mY/(\alpha_D m_X)$.

\begin{figure}
\centering
\includegraphics[width=8cm,height=8cm]{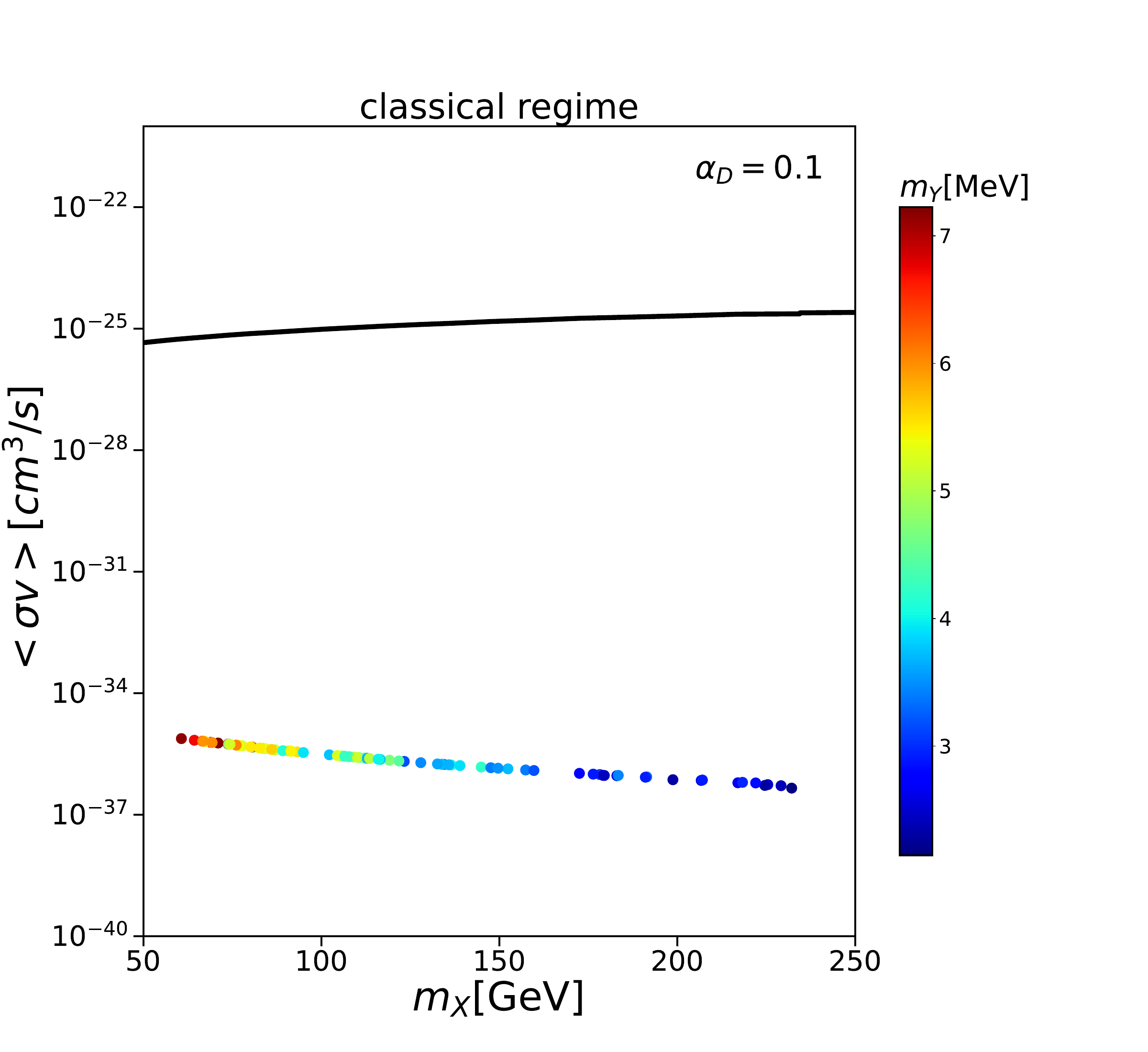}
\includegraphics[width=8cm,height=8cm]{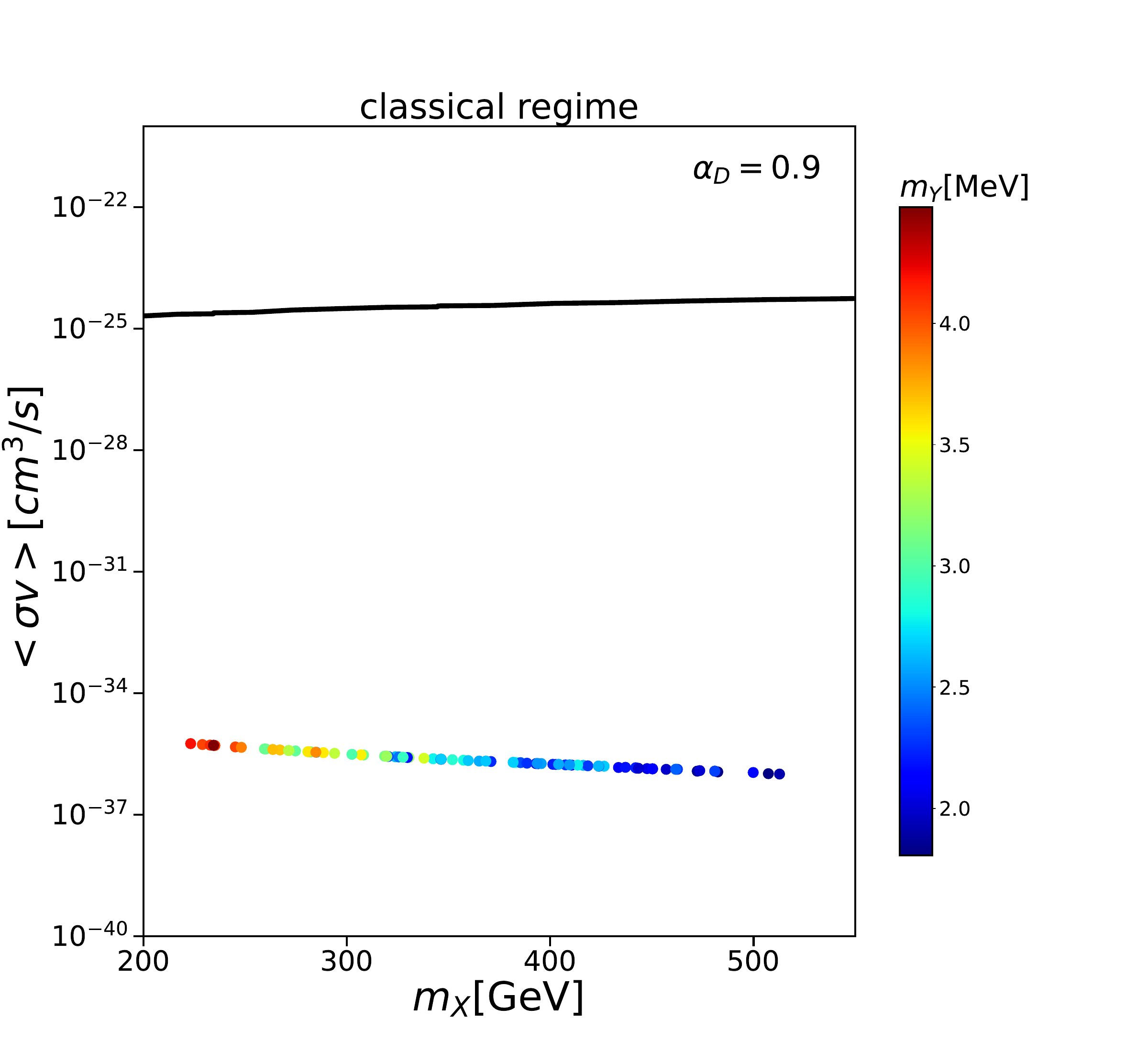}\\
\includegraphics[width=8cm,height=8cm]{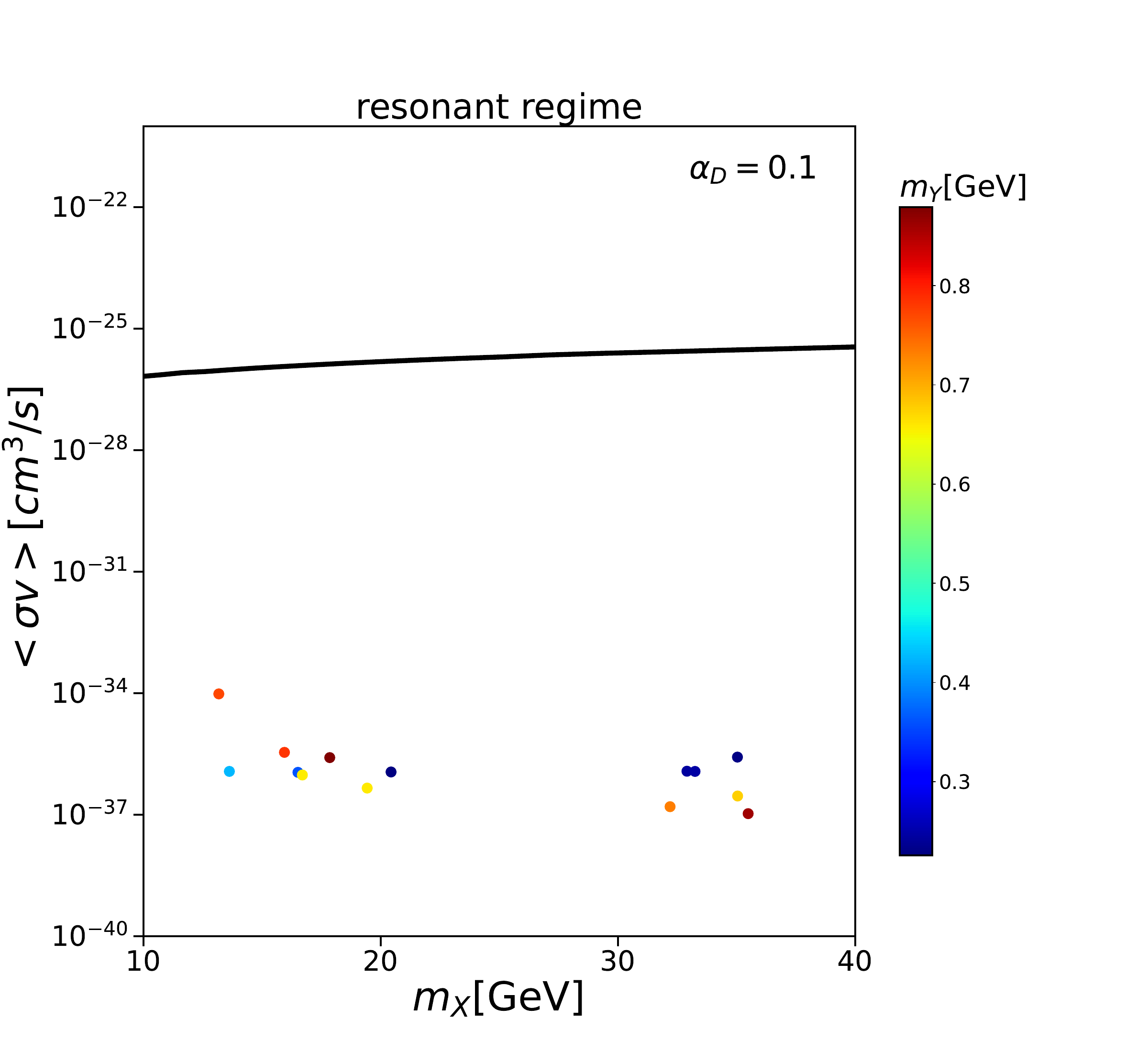}
\includegraphics[width=8cm,height=8cm]{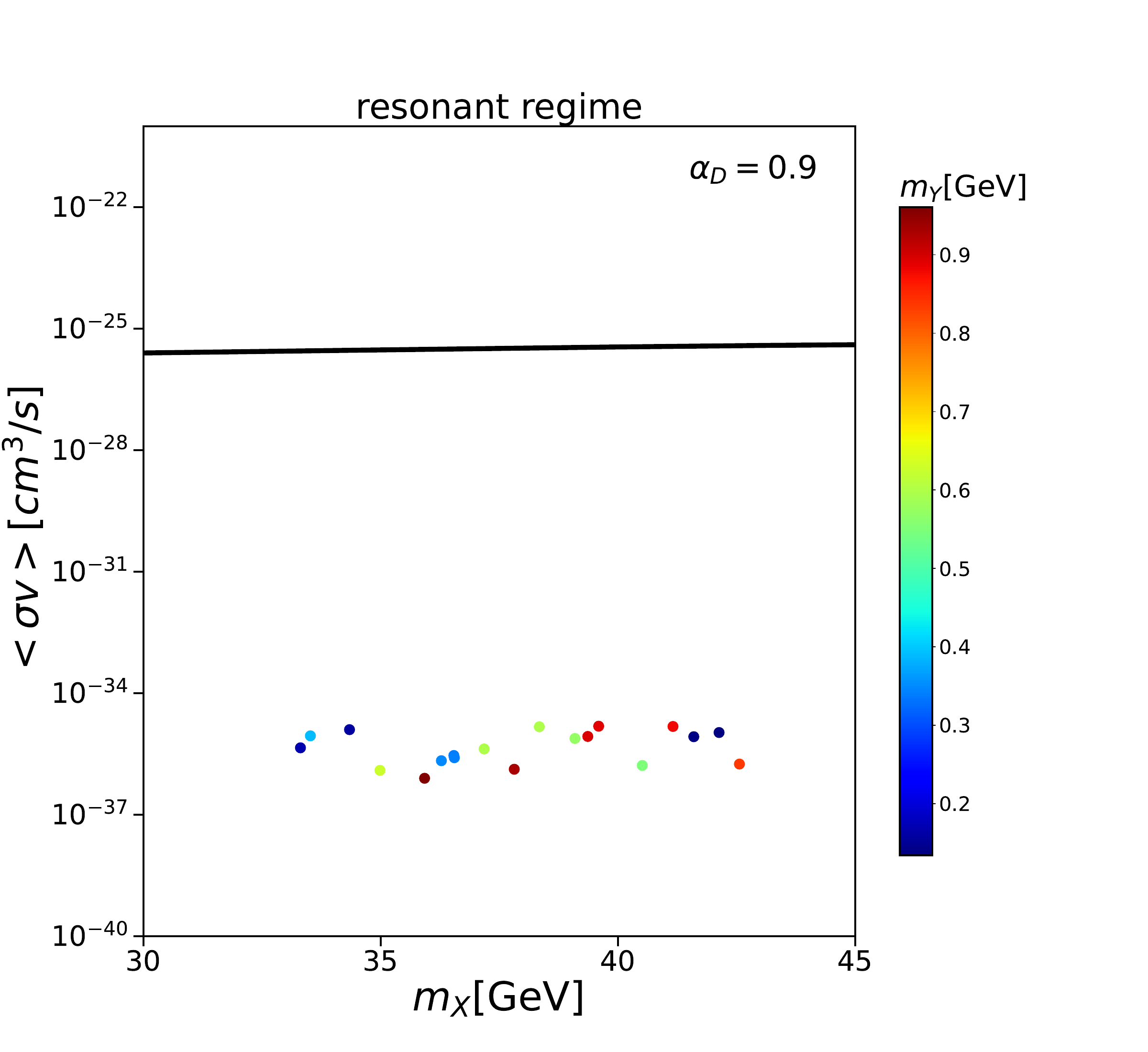}\\
\centering
\caption{The CMB (ann) constraint on $(\sigma_{\rm{ann}}v)_{e\bar{e}}$ (in black) in the classical (top) and resonant (bottom) regime for $\lambda_{D}=0.1$ (left)  and $0.9$ (right) extracted from figs.\ref{Classf}-\ref{Resf} respectively, 
where the values of millicharge $\epsilon$ are fixed by the observed DM relic density in fig.\ref{relic}.}
\label{cmbann}
\end{figure}

Fig.\ref{cmbann} presents the CMB (ann) constraint on $(\sigma_{\rm{ann}}v)_{e\bar{e}}$ (in black) in the classical (top) and resonant (bottom) regime for $\lambda_{D}=0.1$ (left) and $0.9$ (right) extracted from figs.\ref{Classf}-\ref{Resf}  respectively,
where the values of millicharge $\epsilon$ are fixed by the observed DM relic density in fig.\ref{relic}.
This figure clearly illustrates that the CMB (ann) constraint is satisfied either in the classical or resonant regime, 
despite large Sommerfeld effect on the DM annihilation cross section taking place.
Replacing $e\bar{e}$ by the other annihilation final states, similar CMB constraints are expected. 
Take the $b\bar{b}$ for example. While the value of $(\sigma_{\rm{ann}}v)_{b\bar{b}}$ is about the color factor larger than that of $(\sigma_{\rm{ann}}v)_{e\bar{e}}$ with respect to an explicit DM mass, 
the CMB upper bound \cite{Slatyer:2015jla} on the former is also a few times larger than the later.
Therefore, the two resulting exclusion limits are comparable with each other.

\begin{figure}
\centering
\includegraphics[width=8cm,height=8cm]{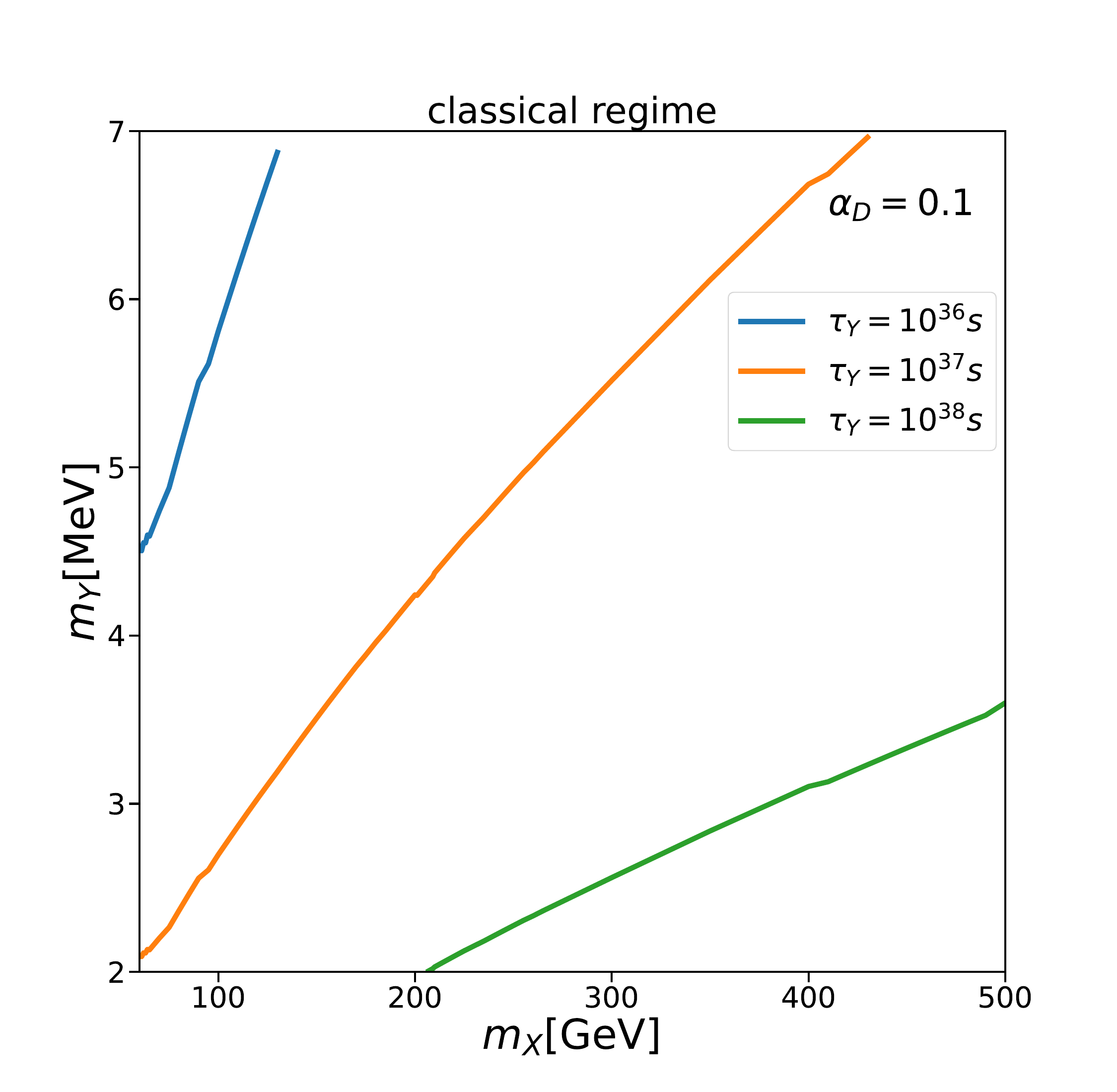}
\includegraphics[width=8cm,height=8cm]{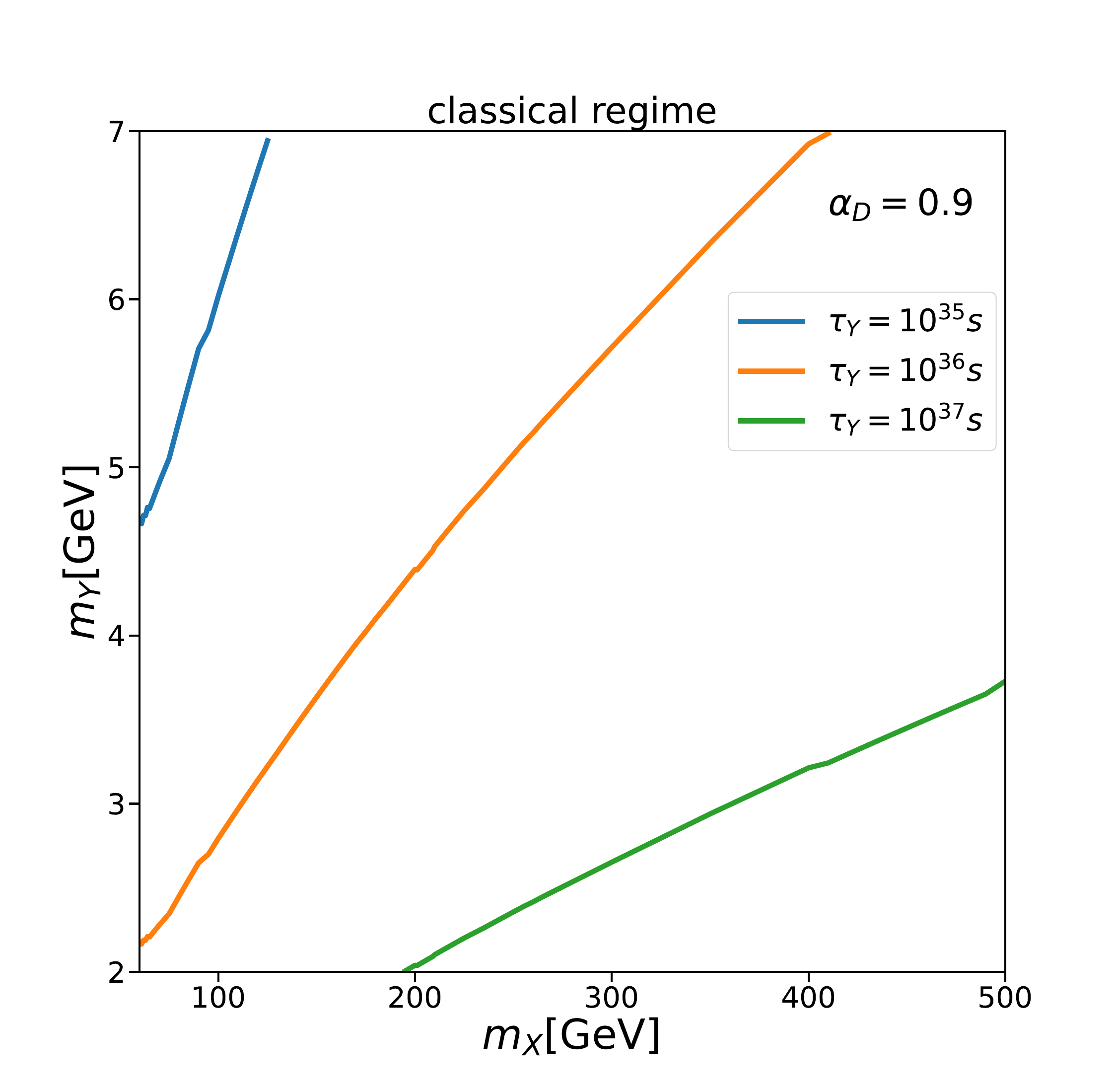}\\
\includegraphics[width=8cm,height=8cm]{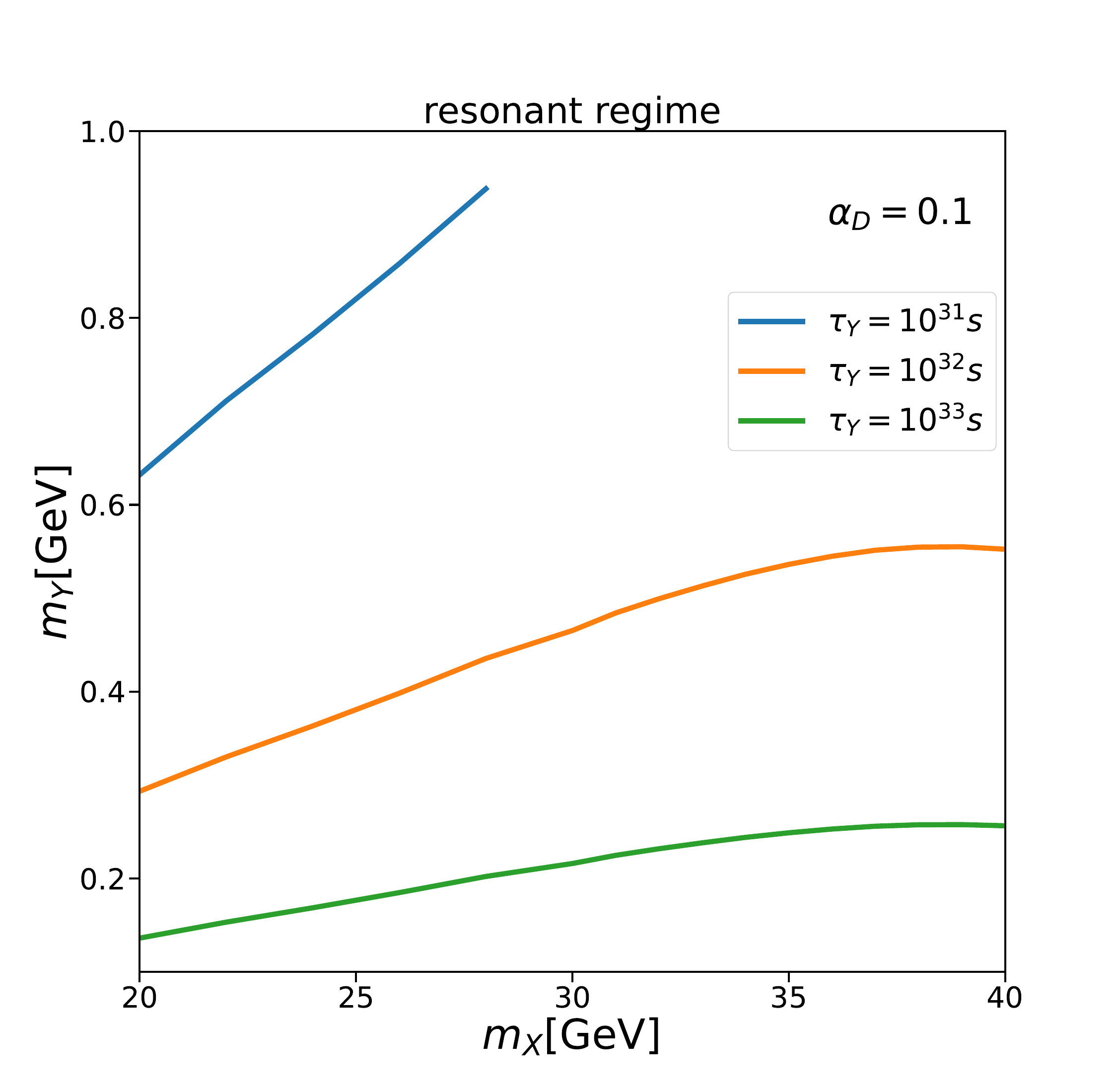}
\includegraphics[width=8cm,height=8cm]{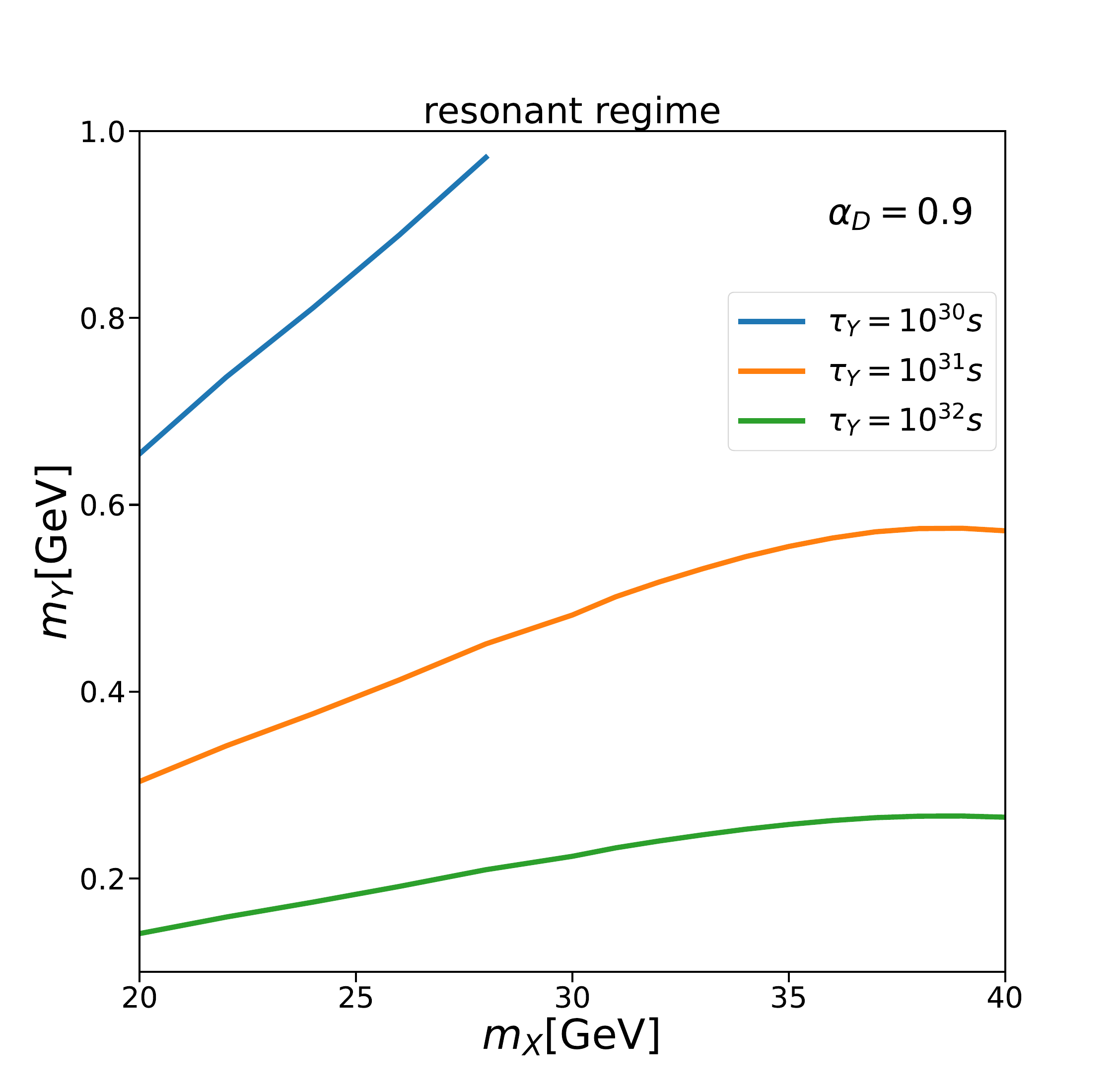}
\centering
\caption{Contours of $\tau_{Y}$ in the classical (top) and resonant (bottom) regime for $\lambda_{D}=0.1$ (left) and $0.9$ (right) respectively, where the values of millicharge $\epsilon$ are fixed by the observed DM relic density shown in fig.\ref{relic}.}
\label{lifetime}
\end{figure}

\subsubsection{Dark matter scattering off baryons}
Apart from constraining the DM annihilations into SM particles,  
CMB data also sets an upper limit on DM scattering off baryons.
For the millicharged DM the scattering cross section scales as $\propto v^{-4}$,
which gives \cite{Dvorkin:2013cea}
\begin{equation}{\label{scat}}
\epsilon< 1.8\times 10^{-6}\left(\frac{m_{X}}{\rm{GeV}}\right)^{1/2}.
\end{equation}
Referring to the plot of DM relic abundance in fig.\ref{relic},
this limit is easily evaded either in the classical or resonant regime.

\subsection{$X/\gamma$-ray}
Unlike the DM being absolutely stable, the DM force mediator radioactively decays to two photons with a long lifetime.
So $X/\gamma$-ray observations can place a constraint on the $Y$ lifetime $\tau_{Y}$.
Fig.\ref{lifetime} presents contours of $\tau_{Y}$ in the classical (top) and resonant (bottom) regime for $\lambda_{D}=0.1$ (left) and $0.9$ (right) respectively,
where we have used
\begin{equation}{\label{dw}}
\tau^{-1}_{Y}\approx
\frac{\epsilon^{4}\alpha^{2}\lambda^{2}_{D}}{144\pi^{3}}\frac{m^{3}_{Y}}{m^{2}_{X}},
\end{equation}
with respect to the effective coupling in eq.(\ref{eff}),
and the values of millicharge $\epsilon$ are fixed by the observed DM relic density as shown in fig.\ref{relic}.
This figure shows that the values of $\tau_{Y}$ range from  $\sim 10^{35}$ to $\sim 10^{38}$ ($\sim 10^{30}$ to $\sim 10^{33}$) sec in the classical (resonant) regime.
Within these lifetime ranges, we impose the following $\gamma/X$-ray constraint \cite{Essig:2013goa,Boddy:2015efa,Riemer-Sorensen:2015kqa}
\begin{equation}{\label{decayc}}
\tau_{Y}\geq 10^{28} \rm{sec} \times\frac{\Omega_{Y}h^{2}}{0.12}.
\end{equation}
Referring to the tiny values of $\Omega_{Y}h^{2}$ as shown by fig.\ref{Yrelic}, the constraint in eq.(\ref{decayc}) is trivially satisfied either in the classical or resonant regime.

\subsection{Supernova 1987A}
Obviously the DM with $m_{X}\sim 1-100$ GeV is hardly produced in Supernova 1987A.
Naively, the lighter DM force mediator with $m_{Y}\sim 1-10$ MeV can be produced through Primakoff process $\gamma\rightarrow Y$ and/or photon inverse decay process $\gamma\gamma\rightarrow Y$.
However, the effective coupling of $Y$ to photons in eq.(\ref{eff}) makes $Y$ production rate scale as $\sim \epsilon^{4}$.
The millicharge induced suppression on the $Y$ production rate plays a similar role of Boltzmann suppression on the $X$ production rate.
Compared to $c_{\gamma}\sim 10^{-24}$ GeV$^{-1}$ in eq.(\ref{eff}),
the Supernova 1987A  limits \cite{Jaeckel:2017tud,Chang:2018rso, Lucente:2020whw},
which constrain axion-like coupling in $g_{\gamma}\phi F\tilde{F}$ as $g_{\gamma}\leq (10^{-11}-10^{-9})$ GeV$^{-1}$ in the axion mass range of $m_{\phi}\sim 1-100$ MeV, are no longer effective. 
If a more concrete analysis on this point is needed, these constraints have to be revised by taking into account the difference between $c_\gamma$ and $g_{\gamma}$.

\subsection{Out-of-equilibrium condition}
The out-of-equilibrium constraint on the DM sector requires $n_{X}\left<(\sigma_{\rm{ann}}v)_{YY}\right>\leq H$ \cite{Chu:2011be,Bernal:2015ova} during the DM freeze-in process, with $H$ the Hubble rate. 
Since $H$ ($n_X$) decreases (increases) as the temperature $T$ of SM thermal bath decreases down to $\sim m_{X}$, 
we can recast this condition as
\begin{eqnarray}\label{oe}
\left<(\sigma_{\rm{ann}}v)_{YY}\right> \leq \frac{H(T)}{n_{X}(T)}\bigg|_{T\approx m_{X}} \approx \frac{\sqrt{\frac{\pi^{2}}{90}}g^{1/2}_{*}m_{X}^{3}}{\Omega_{X}h^{2}\rho_{\rm{crit}}M_{P}}, 
\end{eqnarray}
where $\rho_{\rm{crit}}\approx 8.1h^{-2}\times 10^{-11}\rm{eV}^{4}$ is the critical energy density, $g_{*}=106.5$ is the number of degrees of freedom in the radiation energy density,  
and $M_P$ is the reduced Planck mass.

As mentioned above,  the Sommerfeld effect on the DM annihilation cross section in eq.(\ref{oe}) can be safely neglected during the DM freeze-in process. 
Therefore, using the result in \cite{Han:2017qkr} we obtain
\begin{eqnarray}\label{annY}
(\sigma_{\rm{ann}}v)_{YY}\approx \frac{24\pi\alpha^{2}_{D}}{m^{2}_{X}}v^{2}\approx \frac{24\pi\alpha^{2}_{D}}{m^{2}_{X}}
\end{eqnarray}
where the mass limit $m_{Y}<<m_{X}$ has been used.
Comparing eq.(\ref{annY}) to the right-hand side of  eq.(\ref{oe}) suggests that $m_{X}$ should be larger than $\sim$ keV scale for $\alpha_{D}\sim 1$. 
In this sense, the out-of-equilibrium constraint is trivially satisfied either in the classical or resonant regime.

\section{Conclusion}
\label{con}
Motived by the challenge of WIMP-like DM as the solution to the small-scale problem,
in this work we have proposed a self-interacting FIDM to resolve this issue. 
The DM sector, built upon the SM $U(1)_Y$ portal,
is composed of two components - the DM $X$ and its force mediator $Y$.
We firstly use current small-scale data to identify both the classical and resonant regime resolving  the small-scale problem. 
Then, according to the portal interaction being $X$-SM type with the millicharge as the feeble coupling, 
we use the observed DM relic density due to both the $Z$ and $\gamma$ induced freeze-in processes to fix the remaining model parameter - the DM millicharge $\epsilon$.
Meanwhile, we have verified that the DM force mediator relic abundance is negligible as a result of $\epsilon^{4}$-suppression on its freeze-in production.
 
Within the classical and resonant regime uncovered, 
we have considered various constraints including the CMB constraints on the DM annihilations and the DM scattering off baryons,
the $X/\gamma$-ray constraint on the DM force mediator as it has a lifetime bigger than the age of Universe,
the Supernova 1987A constraint on the DM force mediator as $m_Y$ is of order MeV scale in the classical regime, 
and the out-of-equilibrium constraint on the entire DM sector during the DM freeze-in.
We have verified that the CMB (ann) constraint on the DM annihilations is satisfied despite large Sommerfeld effects taking place, 
whereas the other constraints are trivially accommodated due to various millicharged induced suppressions that have been explained in details. 

We leave tests of this self-interacting FIDM model for future work. 
There are two promising directions. 
A direction is to compare this model prediction on the DM scattering cross section to future small-scale data with respect to $\left<v\right>\sim 100$ km/s.
Another direction is to constrain this DM model using future Lyman-$\alpha$ or 21-cm data.
Because the Sommerfeld enhanced DM annihilation $X\bar{X}\rightarrow e\bar{e}$ contributes to a new energy injection into the intergalactic medium,  
which may lead to observable deviations from baseline ionization history at certain redshift regions.

\section*{Acknowledgments}
S. Xu is supported in part by the National Natural Science Foundation of China (no. 12381240134) and Natural Science Foundation of Henan Province (no. 232300421358).


\begin{thebibliography}{99}

\bibitem{Bernal:2017kxu}
N.~Bernal, M.~Heikinheimo, T.~Tenkanen, K.~Tuominen and V.~Vaskonen,
Int. J. Mod. Phys. A \textbf{32}, no.27, 1730023 (2017),
[arXiv:1706.07442 [hep-ph]].


\bibitem{Hall:2009bx}
L.~J.~Hall, K.~Jedamzik, J.~March-Russell and S.~M.~West,
JHEP \textbf{03} (2010), 080,
[arXiv:0911.1120 [hep-ph]].


\bibitem{Spergel:1999mh}
D.~N.~Spergel and P.~J.~Steinhardt,
Phys. Rev. Lett. \textbf{84} (2000), 3760-3763,
[arXiv:astro-ph/9909386 [astro-ph]].


\bibitem{Rocha:2012jg}
M.~Rocha, A.~H.~G.~Peter, J.~S.~Bullock, M.~Kaplinghat, S.~Garrison-Kimmel, J.~Onorbe and L.~A.~Moustakas,
Mon. Not. Roy. Astron. Soc. \textbf{430}, 81-104 (2013),
[arXiv:1208.3025 [astro-ph.CO]].


\bibitem{Peter:2012jh}
A.~H.~G.~Peter, M.~Rocha, J.~S.~Bullock and M.~Kaplinghat,
Mon. Not. Roy. Astron. Soc. \textbf{430}, 105 (2013),
[arXiv:1208.3026 [astro-ph.CO]].

\bibitem{Zavala:2012us}
J.~Zavala, M.~Vogelsberger and M.~G.~Walker,
Mon. Not. Roy. Astron. Soc. \textbf{431}, L20-L24 (2013),
[arXiv:1211.6426 [astro-ph.CO]].

\bibitem{Elbert:2014bma}
O.~D.~Elbert, J.~S.~Bullock, S.~Garrison-Kimmel, M.~Rocha, J.~O\~norbe and A.~H.~G.~Peter,
Mon. Not. Roy. Astron. Soc. \textbf{453}, no.1, 29-37 (2015),
[arXiv:1412.1477 [astro-ph.GA]].

\bibitem{Dave:2000ar}
R.~Dave, D.~N.~Spergel, P.~J.~Steinhardt and B.~D.~Wandelt,
Astrophys. J. \textbf{547} (2001), 574-589,
[arXiv:astro-ph/0006218 [astro-ph]].

\bibitem{1201.5892}
M.~Vogelsberger, J.~Zavala and A.~Loeb,
Mon.\ Not.\ Roy.\ Astron.\ Soc.\  {\bf 423}, 3740 (2012),
arXiv:1201.5892 [astro-ph.CO].

\bibitem{1208.3025}
M.~Rocha et al.,
Mon.\ Not.\ Roy.\ Astron.\ Soc.\  {\bf 430}, 81 (2013),
arXiv:1208.3025 [astro-ph.CO].

\bibitem{1412.1477}
O.~D.~Elbert, J.~S.~Bullock, S.~Garrison-Kimmel, M.~Rocha, J.~OÃ±orbe and A.~H.~G.~Peter,
Mon.\ Not.\ Roy.\ Astron.\ Soc.\  {\bf 453}, 29 (2015),
arXiv:1412.1477 [astro-ph.GA].

\bibitem{Oh:2010ea}
S.~H.~Oh, W.~J.~G.~de Blok, E.~Brinks, F.~Walter and R.~C.~Kennicutt, Jr,
Astron. J. \textbf{141}, 193 (2011).

\bibitem{KuziodeNaray:2007qi}
R.~Kuzio de Naray, S.~S.~McGaugh and W.~J.~G.~de Blok,
Astrophys. J. \textbf{676}, 920-943 (2008).


\bibitem{Kaplinghat:2015aga}
M.~Kaplinghat, S.~Tulin and H.~B.~Yu,
Phys. Rev. Lett. \textbf{116} (2016) no.4, 041302,
[arXiv:1508.03339 [astro-ph.CO]].



\bibitem{Robertson:2016xjh}
A.~Robertson, R.~Massey and V.~Eke,
Mon. Not. Roy. Astron. Soc. \textbf{465} (2017) no.1, 569-587,
[arXiv:1605.04307 [astro-ph.CO]].

\bibitem{Elbert:2016dbb}
O.~D.~Elbert, J.~S.~Bullock, M.~Kaplinghat, S.~Garrison-Kimmel, A.~S.~Graus and M.~Rocha,
Astrophys. J. \textbf{853} (2018) no.2, 109,
[arXiv:1609.08626 [astro-ph.GA]].

\bibitem{Bondarenko:2017rfu}
K.~Bondarenko, A.~Boyarsky, T.~Bringmann and A.~Sokolenko,
JCAP \textbf{04} (2018), 049,
[arXiv:1712.06602 [astro-ph.CO]].

\bibitem{Harvey:2018uwf}
D.~Harvey, A.~Robertson, R.~Massey and I.~G.~McCarthy,
Mon. Not. Roy. Astron. Soc. \textbf{488} (2019) no.2, 1572-1579,
[arXiv:1812.06981 [astro-ph.CO]].

\bibitem{Sagunski:2020spe}
L.~Sagunski, S.~Gad-Nasr, B.~Colquhoun, A.~Robertson and S.~Tulin,
JCAP \textbf{01}, 024 (2021).

\bibitem{Newman:2015kzv}
A.~B.~Newman, R.~S.~Ellis and T.~Treu,
Astrophys. J. \textbf{814}, no.1, 26 (2015).

\bibitem{Newman:2012nw}
A.~B.~Newman, T.~Treu, R.~S.~Ellis and D.~J.~Sand,
Astrophys. J. \textbf{765}, 25 (2013).

\bibitem{Newman:2012nv}
A.~B.~Newman, T.~Treu, R.~S.~Ellis, D.~J.~Sand, C.~Nipoti, J.~Richard and E.~Jullo,
Astrophys. J. \textbf{765}, 24 (2013).






\bibitem{Tulin:2017ara}
S.~Tulin and H.~B.~Yu,
Phys. Rept. \textbf{730}, 1-57 (2018),
[arXiv:1705.02358 [hep-ph]].

\bibitem{Buckley:2009in}
M.~R.~Buckley and P.~J.~Fox,
Phys. Rev. D \textbf{81} (2010), 083522,
[arXiv:0911.3898 [hep-ph]].


\bibitem{Tulin:2012wi}
S.~Tulin, H.~B.~Yu and K.~M.~Zurek,
Phys. Rev. Lett. \textbf{110} (2013) no.11, 111301,
[arXiv:1210.0900 [hep-ph]].

\bibitem{Tulin:2013teo}
S.~Tulin, H.~B.~Yu and K.~M.~Zurek,
Phys. Rev. D \textbf{87}, no.11, 115007 (2013),
[arXiv:1302.3898 [hep-ph]].

\bibitem{Colquhoun:2020adl}
B.~Colquhoun, S.~Heeba, F.~Kahlhoefer, L.~Sagunski and S.~Tulin,
Phys. Rev. D \textbf{103}, no.3, 035006 (2021),
[arXiv:2011.04679 [hep-ph]].


\bibitem{Hessler:2016kwm}
A.~G.~Hessler, A.~Ibarra, E.~Molinaro and S.~Vogl,
JHEP \textbf{01}, 100 (2017),
[arXiv:1611.09540 [hep-ph]].


\bibitem{Ghosh:2017vhe}
A.~Ghosh, T.~Mondal and B.~Mukhopadhyaya,
JHEP \textbf{12}, 136 (2017),
[arXiv:1706.06815 [hep-ph]].

\bibitem{Calibbi:2018fqf}
L.~Calibbi, L.~Lopez-Honorez, S.~Lowette and A.~Mariotti,
JHEP \textbf{09}, 037 (2018),
[arXiv:1805.04423 [hep-ph]].

\bibitem{Belanger:2018sti}
G.~B\'elanger, N.~Desai, A.~Goudelis, J.~Harz, A.~Lessa, J.~M.~No, A.~Pukhov, S.~Sekmen, D.~Sengupta and B.~Zaldivar, \textit{et al.}
JHEP \textbf{02}, 186 (2019),
[arXiv:1811.05478 [hep-ph]].



\bibitem{Sommerfeld}
A.~Sommerfeld, Annalen der Physik 403, 207 (1931).



\bibitem{Holdom:1985ag}
B.~Holdom,
Phys. Lett. B \textbf{166}, 196-198 (1986).

\bibitem{Feldman:2007wj}
D.~Feldman, Z.~Liu and P.~Nath,
Phys. Rev. D \textbf{75}, 115001 (2007),
[arXiv:hep-ph/0702123 [hep-ph]].




\bibitem{Davidson:2000hf}
S.~Davidson, S.~Hannestad and G.~Raffelt,
JHEP \textbf{05}, 003 (2000),
[arXiv:hep-ph/0001179 [hep-ph]].


\bibitem{Chuzhoy:2008zy}
L.~Chuzhoy and E.~W.~Kolb,
JCAP \textbf{07}, 014 (2009),
[arXiv:0809.0436 [astro-ph]].


\bibitem{McDermott:2010pa}
S.~D.~McDermott, H.~B.~Yu and K.~M.~Zurek,
Phys. Rev. D \textbf{83}, 063509 (2011),
[arXiv:1011.2907 [hep-ph]].


\bibitem{Dolgov:2013una}
A.~D.~Dolgov, S.~L.~Dubovsky, G.~I.~Rubtsov and I.~I.~Tkachev,
Phys. Rev. D \textbf{88}, no.11, 117701 (2013),
[arXiv:1310.2376 [hep-ph]].


\bibitem{Dvorkin:2013cea}
C.~Dvorkin, K.~Blum and M.~Kamionkowski,
Phys. Rev. D \textbf{89}, no.2, 023519 (2014),
[arXiv:1311.2937 [astro-ph.CO]].




\bibitem{Belanger:2018ccd}
G.~B\'elanger, F.~Boudjema, A.~Goudelis, A.~Pukhov and B.~Zaldivar,
Comput. Phys. Commun. \textbf{231}, 173-186 (2018),
[arXiv:1801.03509 [hep-ph]].

\bibitem{Planck:2018vyg}
N.~Aghanim \textit{et al.} [Planck],
Astron. Astrophys. \textbf{641}, A6 (2020)
[erratum: Astron. Astrophys. \textbf{652}, C4 (2021)],
[arXiv:1807.06209 [astro-ph.CO]].


\bibitem{Du:2021jcj}
Y.~Du, F.~Huang, H.~L.~Li, Y.~Z.~Li and J.~H.~Yu,
JCAP \textbf{04}, no.04, 012 (2022),
[arXiv:2111.01267 [hep-ph]].









\bibitem{Slatyer:2015jla}
T.~R.~Slatyer,
Phys. Rev. D \textbf{93}, no.2, 023527 (2016),
[arXiv:1506.03811 [hep-ph]].


\bibitem{Feng:2010zp}
J.~L.~Feng, M.~Kaplinghat and H.~B.~Yu,
Phys. Rev. D \textbf{82}, 083525 (2010),
[arXiv:1005.4678 [hep-ph]].

\bibitem{Iengo:2009ni}
R.~Iengo, 
JHEP \textbf{05}, 024 (2009),
[arXiv:0902.0688 [hep-ph]].

\bibitem{Cassel:2009wt}
S.~Cassel,
J. Phys. G \textbf{37}, 105009 (2010),
[arXiv:0903.5307 [hep-ph]].

\bibitem{Slatyer:2009vg}
T.~R.~Slatyer,
JCAP \textbf{02}, 028 (2010),
[arXiv:0910.5713 [hep-ph]].



\bibitem{Essig:2013goa}
R.~Essig, E.~Kuflik, S.~D.~McDermott, T.~Volansky and K.~M.~Zurek,
JHEP \textbf{11}, 193 (2013),
[arXiv:1309.4091 [hep-ph]].

\bibitem{Boddy:2015efa}
K.~K.~Boddy and J.~Kumar,
Phys. Rev. D \textbf{92}, no.2, 023533 (2015),
[arXiv:1504.04024 [astro-ph.CO]].


\bibitem{Riemer-Sorensen:2015kqa}
S.~Riemer-S\o{}rensen, D.~Wik, G.~Madejski, S.~Molendi, F.~Gastaldello, F.~A.~Harrison, W.~W.~Craig, C.~J.~Hailey, S.~E.~Boggs and F.~E.~Christensen, \textit{et al.}
Astrophys. J. \textbf{810}, no.1, 48 (2015),
[arXiv:1507.01378 [astro-ph.CO]].




\bibitem{Jaeckel:2017tud}
J.~Jaeckel, P.~C.~Malta and J.~Redondo,
Phys. Rev. D \textbf{98}, no.5, 055032 (2018),
[arXiv:1702.02964 [hep-ph]].

\bibitem{Chang:2018rso}
J.~H.~Chang, R.~Essig and S.~D.~McDermott,
JHEP \textbf{09}, 051 (2018),
[arXiv:1803.00993 [hep-ph]].

\bibitem{Lucente:2020whw}
G.~Lucente, P.~Carenza, T.~Fischer, M.~Giannotti and A.~Mirizzi,
JCAP \textbf{12}, 008 (2020),
[arXiv:2008.04918 [hep-ph]].




\bibitem{Chu:2011be}
X.~Chu, T.~Hambye and M.~H.~G.~Tytgat,
JCAP \textbf{05}, 034 (2012),
[arXiv:1112.0493 [hep-ph]].

\bibitem{Bernal:2015ova}
N.~Bernal, X.~Chu, C.~Garcia-Cely, T.~Hambye and B.~Zaldivar,
JCAP \textbf{03}, 018 (2016)
[arXiv:1510.08063 [hep-ph]].

\bibitem{Han:2017qkr}
H.~Han, H.~Wu and S.~Zheng,
Chin. Phys. C \textbf{43}, no.4, 043103 (2019),
[arXiv:1711.10097 [hep-ph]].




\end{thebibliography}
\end{document}